\documentclass[12pt, draftclsnofoot, onecolumn]{IEEEtran}

\usepackage{graphicx}
\usepackage{amsmath,epsfig}
\usepackage{amssymb}
\usepackage{upgreek}
\usepackage{bm}
\usepackage{authblk}
\usepackage{setspace}
\usepackage{termcal}
\usepackage{caption}
\usepackage{subcaption}
\usepackage{siunitx}
\usepackage{fancyhdr}
\usepackage{acro}
\usepackage{bbm}
\usepackage{xcolor}
\usepackage{mathtools}

\usepackage[pagewise]{lineno}

\DeclareAcronym{awgn}{short = AWGN, long = additive white Gaussian noise}
\DeclareAcronym{ici}{short = ICI, long = intercarrier interference}
\DeclareAcronym{sinr}{short = SINR, long = signal-to-interference-and-noise ratio}
\DeclareAcronym{snr}{short = SNR, long = signal-to-noise ratio}
\DeclareAcronym{ofdm}{short = OFDM, long = orthogonal frequency-division multiplexing}
\DeclareAcronym{mimo}{short = MIMO, long = multiple-input multiple-output}
\DeclareAcronym{cfo}{short = CFO, long =  carrier frequency offset}
\DeclareAcronym{cpe}{short = CPE, long = common phase error}
\DeclareAcronym{lte}{short = LTE, long = long term evolution}
\DeclareAcronym{nr}{short = NR, long = new radio}
\DeclareAcronym{crlb}{short = CRLB, long = Cramer-Rao Lower Bound}
\DeclareAcronym{acf}{short = ACF, long = autocorrelation function}
\DeclareAcronym{zzb}{short = ZZB, long = Ziv-Zakai Bound}
\DeclareAcronym{mmse}{short = MMSE, long = minimum mean square error}
\DeclareAcronym{lmmse}{short = LMMSE, long = linear minimum mean square error}
\DeclareAcronym{rmse}{short = RMSE, long = root mean square error}
\DeclareAcronym{mrc}{short = MRC, long = maximum-ratio combining}
\DeclareAcronym{toa}{short = TOA, long = time-of-arrival}
\DeclareAcronym{pdf}{short = PDF, long = probability density function}
\DeclareAcronym{cdf}{short = CDF, long = cumulative distribution function}
\DeclareAcronym{dft}{short = DFT, long = discrete Fourier transform}
\DeclareAcronym{tdoa}{short = TDOA, long = time-difference-of-arrival}
\DeclareAcronym{rtt}{short = RTT, long = round-trip-time}

\newcommand{\bpshz}[1]{\SI{#1}{bit\per\second\per\hertz}}

\begin{document}
	\bstctlcite{IEEEmax3beforeetal}
	
	\title{Purposeful Co-Design of OFDM Signals for Ranging and Communications}
	
	\author{Andrew Graff \IEEEmembership{Student Member, IEEE} and Todd E. Humphreys \IEEEmembership{Member, IEEE}
		\thanks{A. Graff is with the Department of Electrical and Computer Engineering, The University of Texas at Austin, Austin, TX 78712, USA \mbox{(e-mail: andrewgraff@utexas.edu)}.}
		\thanks{T. Humphreys is with the Department of Aerospace Engineering and Engineering Mechanics, The University of Texas at Austin, Austin, TX 78712, USA \mbox{(e-mail: todd.humphreys@utexas.edu)}.}
	}
	
	\maketitle
	
	\newif\ifpreprint
	\preprinttrue
	
	\ifpreprint
	
	\pagestyle{plain}
	\thispagestyle{fancy}  
	\fancyhf{} 
	\renewcommand{\headrulewidth}{0pt}
	\lfoot{\footnotesize \bf Copyright \copyright~2024 by Andrew Graff, \\
		and Todd E. Humphreys}
	\rfoot{\parbox{0.6\linewidth}{\footnotesize \bf This version of the article has been accepted for publication, after peer review, but is not the Version of Record and does not reflect post-acceptance improvements, or any corrections. The Version of Record is available online at: http://dx.doi.org/10.1186/s13634-024-01110-w.}}
	\else
	
	\thispagestyle{empty}
	\pagestyle{empty}
	
	\fi
	
	\begin{abstract}
		This paper analyzes the fundamental trade-offs that occur in the co-design of pilot resource allocations in orthogonal frequency-division multiplexing signals for both ranging (via time-of-arrival estimation) and communications. These trade-offs are quantified through the Shannon capacity bound, probability of outage, and the Ziv-Zakai bound on range estimation variance. Bounds are derived for signals experiencing frequency-selective Rayleigh block fading, accounting for the impact of limited channel knowledge and multi-antenna reception. Uncompensated carrier frequency offset and phase errors are also factored into the capacity bounds. Analysis based on the derived bounds demonstrates how Pareto-optimal design choices can be made to optimize the communication throughput, probability of outage, and ranging variance. Different pilot resource allocation strategies are then analyzed, showing how Pareto-optimal design choices change depending on the channel.
	\end{abstract}
	
	
	\section{Introduction}
	
	Today's wireless communication networks are experiencing an ever-growing demand not only for traditional communications but also for positioning, navigation, and timing services, especially accurate user localization. As user equipment (UE) is deployed in increasingly mobile contexts, ranging from pedestrian to automotive to aerospace applications, the next generation of wireless networks will need to meet high demands for precise positioning. \Ac{ofdm}, the basis of the widely-deployed standards 802.11, long term evolution (LTE), and 5G new radio (NR), is by far the most commonly used modulation for broadband wireless networks. \ac{ofdm}-based standards currently include positioning protocols \cite{3GPP,keating2019overview}, but the protocols have a narrowly-limited range of performance options because they are included as add-ons within an overall signal structure designed to prioritize communications metrics, e.g., data rates, latency, and network reliability. Furthermore, the positioning protocols of existing standards only offer positioning to authorized network users, which precludes their use in the context of signals of opportunity with the attendant benefits of anonymity and multi-network positioning \cite{shamaei2018exploiting,shamaei2018lte,kassas2014receding}. Against the backdrop of these significant limitations and in view of the increasing importance of positioning in wireless networks, the current paper considers the fundamental theoretical tradeoffs inherent in co-design of the pilot resource allocations in \ac{ofdm} waveforms for both communications and positioning. It then proceeds to explore the pilot resource designs that optimize communications or positioning while satisfying a threshold requirement for the other.
	
	\begin{figure}[h!]
		\centering
		\includegraphics[width=0.45\textwidth,page=4]{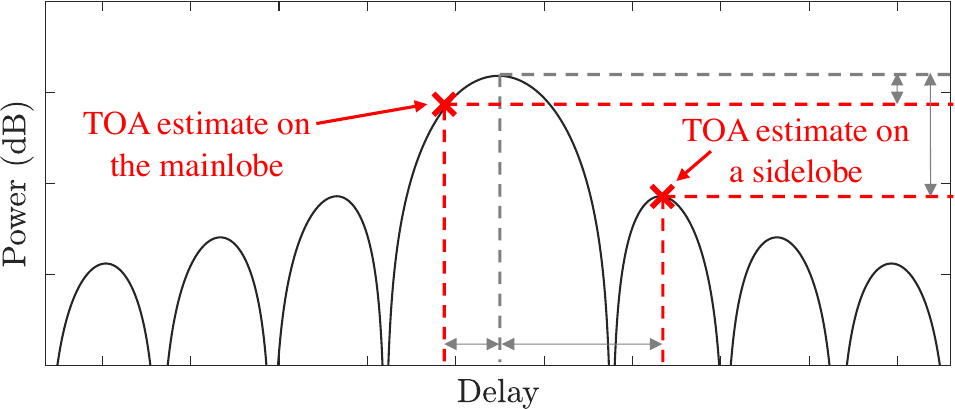}
		\caption{A graphical depiction of an autocorrelation function, marking a \ac{toa} estimate on the mainlobe and a \ac{toa} estimate on the sidelobe. The mainlobe estimate has a smaller difference in power relative to the peak than the sidelobe estimate. The mainlobe estimate also has a smaller error in delay.}
		\label{fig:acf_sidelobe}
	\end{figure}
	
	Consider a scheme in which a UE attempts to determine its position through \acf{toa} estimates of signals arriving from multiple cellular base stations. The simplest approach the UE can take is to correlate its received samples against a local replica signal constructed from the known \ac{ofdm} pilot resources. Through this correlation, the UE can extract a \ac{toa} estimate \cite{berger2010signal} relative to its local clock. This \ac{toa} estimate may then be used as a pseudorange measurement \cite{shamaei2018exploiting}, a \ac{tdoa} measurement if subtracted against measurements from other base stations \cite{dwivedi2021positioning}, or as part of a \ac{rtt} measurement if the base station measures the \ac{toa} in the uplink direction as well \cite{dwivedi2021positioning}. Much like in traditional global navigation satellite system processing, the UE can then determine its location and time relative to the known locations of the base stations and to the network time. However, the requirement to collect \ac{toa} estimates from multiple base stations may not be easily satisfied in wireless networks, which are often designed to serve a given region with a single base station. In this setting, the \acp{snr} of signals collected from more distant base stations may be so low that the associated \ac{toa} estimates experience a thresholding effect where their variance rises dramatically. This thresholding effect has been thoroughly studied in the analysis of fundamental bounds on parameter estimation, such as the Barankin bound \cite{barankin1949,mcaulay1971barankin} and the Ziv-Zakai bound \cite{Ziv1969}, and also in the context of \ac{toa} estimation \cite{zeira1994realizable,nanzer2016bandpass,sahinoglu2008ultra}. The thresholding phenomenon occurs when sidelobes (secondary peaks) in the autocorrelation function become the dominant source of estimation error. Fig.~\ref{fig:acf_sidelobe} depicts estimation errors that may occur along an autocorrelation function. At a sufficiently high \ac{snr}, the variance of the noise in the receiver's correlation output will be much smaller than the ratio between the peak power of the mainlobe and the peak powers of the sidelobes. As a result, the probability of estimation errors occurring on the sidelobes is negligibly low, and errors are instead concentrated on the mainlobe. In this ``mainlobe-dominated regime,'' \ac{toa} accuracy is determined by the shape of the autocorrelation function near the mainlobe's peak, with a sharper peak yielding greater accuracy. This mainlobe-dominated regime is where the \ac{crlb} on the \ac{toa} error variance applies. However at a sufficiently low \ac{snr}, the probability of estimates settling on the sidelobes becomes non-negligible, causing drastic increases in \ac{toa} error variance, as such \ac{toa} estimates are far from the mainlobe peak. The operating conditions under which this occurs will be referred to as the ``sidelobe-dominated regime.'' As the \ac{snr} drops still further, \ac{toa} error variances continue to rise, eventually plateauing in an ``ambiguous regime'' where meaningful estimates cannot be obtained from the signal and the best estimator is that which maximizes the prior belief about the \ac{toa}.
	
	Additional operating regimes may emerge if the UE uses more advanced techniques than simply correlating against known pilots. One such technique is decision-directed estimation, in which the UE first decodes the previously-unknown data resources, reconstructs the \ac{ofdm} signal using the data resource estimates along with the known pilot resources, and correlates against this reconstructed signal to obtain an improved \ac{toa} estimate \cite{mensing2009dd}. Since data resources make up a significant portion of the energy in \ac{ofdm} transmissions, such a decision-directed estimator enjoys noticeable gains in post-correlation \ac{snr} and thus a decrease in \ac{toa} estimation error when the bit error rate is sufficiently low. Decision-directed approaches have seen success in channel and Doppler estimation \cite{Kalyani2007,Shi2005}. A similar maximum-likelihood approach called non-data-aided (NDA) estimation has been used for time delay estimation \cite{masmoudi2017nda}. While decision-directed estimators have been used for positioning \cite{mensing2009dd}, further research is needed to quantify their performance in modern cellular networks. Meanwhile, the current paper's focus will be on schemes where \ac{toa} is estimated only through correlation against known elements of an \ac{ofdm} signal.
	
	This paper explores how the design of \ac{ofdm} pilot resources impacts both ranging precision and communication capacity in various propagation environments. Throughout this paper, ``ranging'' is used synonymously with \ac{toa} estimation at the receiver relative to its local clock. This \ac{toa} estimation precedes and is agnostic to the positioning method used (e.g., pseudorange multilateration, \ac{tdoa}, \ac{rtt}, etc.). To quantify ranging precision, it employs the Ziv-Zakai bound \cite{Ziv1969}, which, unlike the \ac{crlb}, accurately captures estimator errors across all applicable regimes. To quantify communication capacity, Shannon capacity and probability of outage are computed, factoring in impairments due to multipath, block fading, \ac{cfo} estimation error, and \ac{cpe}. The placement and power allocation of pilot resources within an \ac{ofdm} block involves intricate tradeoffs between ranging precision and communication throughput, especially when channel impairments are considered. This paper quantifies these tradeoffs and proposes \ac{ofdm} pilot resource designs that balance performance in ranging, capacity, and outages.
	
	\subsection{Contributions}
	The main contributions of this paper are as follows:
	\begin{itemize}
		\item A derivation of the Shannon capacity bound and probability of outage for generic \ac{ofdm} signals, accounting for channel estimation error, intercarrier interference, and common phase errors,  together with a derivation of the Ziv-Zakai bound on range estimation variance for the same signals. These bounds also account for Rayleigh fading and multipath channels.
		\item A method of co-designing the placement and power allocation of pilot resources in an \ac{ofdm} signal to achieve both ranging and communication performance requirements through the use of Pareto curves plotting both the Shannon capacity and probability of outage against the Ziv-Zakai bound on ranging error variance.
		\item An analysis of how different channel impairments, fading models, and multipath affect both communications and ranging performance, including their effect on Pareto-optimal pilot resource allocations.
	\end{itemize}
	
	\subsection{Prior Work}
	
	Prior work has studied how \ac{ofdm} signals can be used for positioning, but much of this work operates only within existing protocols rather than proposing new signal designs. This is a broad field of work, covering several protocols of interest. \ac{toa} and ranging estimators for LTE signals have been analyzed in \cite{Wang2019,PeralRosado2018,xu2016maximum}. The \ac{crlb} for \ac{toa}/range estimation is derived in both \cite{PeralRosado2018} and \cite{xu2016maximum} to evaluate the performance of their estimators, but this bound is inapt for low \ac{snr} regimes. Outside of OFDM signals, \cite{jativa2018cramer} derives the \ac{crlb} for \ac{toa} estimation that exploits temporal correlation in fading channels but is similarly inapt at low \acp{snr}. A comparison between \ac{ofdm} and pseudonoise-based signals in \cite{Wang2010} demonstrates that \ac{ofdm} signals may provide improved time-based range estimation performance. Many publications have also focused on the field of opportunistic positioning and navigation, utilizing signals from LTE \cite{shamaei2018exploiting}, \cite{shamaei2018lte}, FM \ac{ofdm} \cite{psiaki2019tracking}, and mobile TV \cite{serant2010development,Rabinowitz2005}. While these studies provide valuable insights into the performance capabilities of such estimation and positioning algorithms, they do not address the design of the signals themselves, instead working within existing protocols.
	
	Some work has specifically addressed the design of \ac{ofdm} signals for ranging. Driusso et al. consider signal design and study how the placement of positioning pilots within the LTE framework affect ranging performance by computing the Ziv-Zakai bound in \ac{awgn} \cite{Driusso2015}. However, their bounds do not account for fading effects. Wang et al. provide ranging accuracy bounds for a generic \ac{ofdm} signal model that includes multipath fading but only compute the \ac{crlb}, failing to address the \ac{snr} threshold effect \cite{Wang2014}. The study in \cite{Dun2021} proposes a unique \ac{ofdm} design strategy for selecting a sparse subset of dedicated bands such that \ac{toa} estimation can meet given requirements under multipath propagation environments. The multipath signal modeling is rigorous and the estimation computationally complexity is significantly reduced using the proposed sparse design. However, the criterion used for optimization is the \ac{crlb}, which ignores sidelobes and thresholding effects at low \ac{snr}. Furthermore, the study does not directly address how such a ranging signal would coexist within an \ac{ofdm} system also being used for communications. Another optimization technique is proposed in \cite{Larsen2011}, in which pilots are allocated to optimize for both time-delay and channel estimation. Much like the previous paper, the \ac{crlb} is used in the optimization criterion, limiting this technique's applicability at low \ac{snr}, and communication capacity is not addressed. Karisan et al. take a similar approach \cite{Karisan2011} where the power allocation across pilots is designed to minimize the range estimation \ac{crlb} in the presence of interference. But, like others mentioned, this paper does not address low \ac{snr} thresholding effects or the tradeoffs that such a design would have with a joint communication system. While not specifically addressing \ac{ofdm} design for ranging, the impact of \ac{ofdm} design parameters on sidelobe energy in the signal's autocorrelation function was analyzed in \cite{Ureten2009}. The Ziv-Zakai bound has also been applied to \ac{toa} estimation in multipath fading channels from wideband pulse signals when the receiver has prior knowledge of the channel multipath \cite{gifford2020impact,dardari2009ziv}. A multiband-\ac{ofdm} positioning testbed was developed in \cite{koelemeij2022hybrid}, demonstrating decimeter-level accuracy through real-world \ac{toa}-based positioning and a sparse selection of bands.
	
	The communication capacity of \ac{ofdm} systems has also been extensively studied in prior work. Goldsmith's textbook on wireless communications thoroughly covers the computation of channel capacity and outage in the presence of fading \cite{GoldsmithBook}. Yoo and Goldsmith extend this analysis to \ac{mimo} channels with channel estimation error \cite{GoldsmithPaper}. Tang et al.  analyze the effect of channel estimation error in the presence of Rayleigh fading \cite{Tang1999}. Ohno and Giannakis provide analysis on the \ac{mmse} channel estimation error in \ac{ofdm} systems and its impact on channel capacity in block Rayleigh fading \cite{Ohno2004}, using this work to propose optimal pilots to maximize capacity. The study in \cite{sun2023trade} analyzes the tradeoff between ranging and communications using the Ziv-Zakai bound for millimeter-wave signals as the power allocation is varied. However, the study does not analyze the impact of pilot placement or consider multipath effects, which are more common at sub-\SI{6}{\giga\hertz} frequencies. Zhu et al. \cite{zhu2023information} use the \ac{crlb} to optimize beamforming weights to balance communications rate and sensing, but did not analyze \ac{toa} estimation or pilot resource design. While the capacity of these systems alone has been thoroughly analyzed and some tradeoffs with positioning have been provided, such analysis has not been combined with a ranging variance analysis under multipath fading environments to illuminate the trade-offs between capacity and ranging precision. The current paper is the first to jointly analyze the impact that \ac{ofdm} pilot placements and power allocations have on both communications and ranging while considering multipath fading channels, degradation in communications caused by channel, \ac{cfo}, and \ac{cpe} estimation errors, and thresholding effects in \ac{toa} estimation quantified through the use of the Ziv-Zakai bound.
	
	A previous conference publication \cite{graff2021ofdmAnalysisIonGNSS} by the current paper's authors laid the groundwork for this current paper, providing simplified versions of the Shannon capacity, Ziv-Zakai bound, and analysis on the trade-offs experienced in different propagation environments. The current paper significantly extends the conference paper in the following ways. First, it models carrier frequency offset, intercarrier interference, and common phase errors, quantifies the impact that pilot resource allocations have on their estimation errors, and factors the impact of these errors into capacity. Second, it evaluates more realistic \ac{ofdm} signals spanning multiple symbols, capturing the time-dependent aspects of phase errors in the signal. Third, it analyzes the Ziv-Zakai bound with the optimal detector rather than the suboptimal detector in \cite{graff2021ofdmAnalysisIonGNSS} when the receiver has no channel knowledge. Last, it provides a significantly expanded analysis of the results and further quantifies communications performance through outage probability.
	
	\textbf{Notation:} Column vectors are denoted with lowercase bold, e.g., $\bm{x}$. Matrices are denoted with uppercase bold, e.g., $\bm{X}$. Scalars are denoted without bold, e.g., $x$. The $i$th entry of a vector $\bm{x}$ is denoted $x[i]$. The Euclidean norm is denoted $||\bm{x}||$, and the Frobenius norm is denoted $||\bm{X}||_{\text{F}}$. Real transpose is represented by the superscript $T$ and conjugate transpose by the superscript $H$. Element-wise vector multiplication is represented by $\odot$. Circular convolution is represented by $\circledast$. The floor function is denoted as $\lfloor . \rfloor$. The Q-function is denoted as $Q(.)$. Zero-based indexing is used throughout the paper. A superscript $\; \tilde{}\;$ is used throughout this paper to indicate signals in the frequency domain after the receiver's OFDM \ac{dft} processing.
	
	\section{Signal Model}
	\label{section:sig}
	A user receives an \ac{ofdm} signal that has propagated through a finite-impulse-response (FIR) channel in the presence of \ac{awgn}. The signal model is discrete time and baseband. The \ac{ofdm} system operates with $K$ subcarriers and a cyclic prefix length of $L_{\text{c}}$. Let $(\tilde{s}^{(m)}[k])_{k=0}^{K-1}$ be the complex modulation sequence for the $m$th \ac{ofdm} symbol, where $\sigma_{\tilde{s}^{(m)}}^2[k] \triangleq \mathbb{E}\left[|\tilde{s}^{(m)}[k]|^2\right]$ for $k = 0,1,\; \ldots,K-1$. Assume that for the $m$th \ac{ofdm} symbol, pilot resources are placed at subcarrier indices $\mathcal{I}_{\text{pilot}}^{(m)}$ and data resources at subcarrier  indices $\mathcal{I}_{\text{data}}^{(m)}$ such that these sets are disjoint and exhaustive: $\mathcal{I}_{\text{pilot}}^{(m)} \cap \mathcal{I}_{\text{data}}^{(m)} = \emptyset$ and $\mathcal{I}_{\text{pilot}}^{(m)} \cup \mathcal{I}_{\text{data}}^{(m)} = \{k \in \mathbb{Z} : k \in [0,K-1]\}$. These are modulated onto subcarriers using an inverse DFT, creating a complex time-domain signal with $\bar{K} \triangleq K+L_{\text{c}}$ samples per \ac{ofdm} symbol. Suppose $M$ \ac{ofdm} symbols are transmitted. Then the transmitted signal is
	\begin{align}
		w[n] = \frac{1}{K} \sum_{k=0}^{K-1}{\tilde{s}^{(m)} [k]e^{j2\pi{}\frac{k(q -L_{\text{c}})}{K}}}, 
		\quad m = \lfloor n/\bar{K} \rfloor, \quad q = n \bmod{\bar{K}}, \quad n = 0,1,\ldots,M\bar{K}-1.
		\label{eq:transmit_sig}
	\end{align}
	
	This transmitted signal has a sampling rate of $f_{\text{s}}\; \SI{}{\hertz}$, resulting in a sample period of $T_{\text{s}} = \frac{1}{f_{\text{s}}}\; \SI{}{\second}$ and a subcarrier spacing of $\frac{1}{K T_{\text{s}}}\; \SI{}{\hertz}$. With the cyclic prefix, each \ac{ofdm} symbol has a period of $T_{\text{sym}} \triangleq \bar{K}T_{\text{s}}\; \SI{}{\second}$. The first sample of the $m$th \ac{ofdm} symbol is $w[m\bar{K}]$. The time-domain signal propagates through a fading channel modeled as an order-$L$ FIR filter with $\bar{L} \triangleq L + 1$ complex taps $(h[l])_{l=0}^{L}$, written in vector form as $\bm{h}$. The signal is also subject to noise modeled as \ac{awgn} and expressed as the sequence $(v[n])_{n=0}^{M\bar{K}-1}$ where $v[n]\sim\mathcal{C}\mathcal{N}(0,\sigma_{v}^2)$ for $n = 0,1,\ldots, M\bar{K}-1$. It is assumed that $\bar{L}\leq{}L_{\text{c}}$. Additionally, the signal experiences phase rotations due to \ac{cfo} with frequency $f_{\text{cfo}}$, and phase noise is modeled as a sampled Wiener process $(\phi[n])_{n=0}^{M\bar{K}-1}$. Provisionally assuming zero time delay, the resulting received signal from the transmission of (\ref{eq:transmit_sig}) is given by
	\begin{align}
		x[n] = \exp\left(j(2\pi f_{\text{cfo}} T_{\text{s}} (n-L_{\text{c}}) + \phi[n])\right) \sum_{l=0}^{L}{h[l]w[n-l]} + v[n], \quad n = 0,1,\ldots,M\bar{K}-1.
		\label{eq:time_domain_sig}
	\end{align}
	Without loss of generality, the sample indices in the \ac{cfo} term are shifted by $L_{\text{c}}$ so that the first sample after the cyclic prefix in the first symbol experiences zero phase shift due to \ac{cfo}. The cyclic prefix is assumed to be sufficiently long for both the communications and ranging functions, permitting the receiver to obtain the frequency-domain signal by discarding the cyclic prefix samples and taking the DFT of each symbol. For reference, the standard LTE cyclic prefix length of \SI{4.69}{\micro\second} has an equivalent range of \SI{1406}{\meter} \cite{shamaei2018exploiting}. Define the time-domain samples for the $m$th symbol after discarding the cyclic prefix as $x^{(m)}[i] \triangleq x[m\bar{K} + L_{\text{c}} + i]$ for $i = 0,\; \ldots, K-1$, expressed in vector form as $\bm{x}^{(m)}$. The noise component of these samples is $v^{(m)}[i] \triangleq v[m\bar{K} + L_{\text{c}} + i]$ for $i = 0,\; \ldots, K-1$, expressed in vector form as $\bm{v}^{(m)}$. Let $\bm{F} \in \mathbb{C}^{K \times K}$ be a DFT matrix, normalized such that $||\bm{F}||_{\text{F}} = \sqrt{K}$, and let $\bm{F}_{\bar{L}}$ represent the first $\bar{L}$ columns of $\bm{F}$. The frequency-domain signal for symbol $m$ is $\tilde{\bm{x}}^{(m)} = \bm{F}\bm{x}^{(m)}$. The frequency-domain channel coefficients are denoted $\tilde{\bm{h}} = \sqrt{K}\bm{F}_{\bar{L}}\bm{h}$, and the frequency-domain noise is denoted $\tilde{\bm{v}}^{(m)} = \bm{F} \bm{v}^{(m)}$. Assuming the process $\phi[n]$ is slowly varying and can be approximated as constant over the duration of one \ac{ofdm} symbol, $\phi[n]$ is approximated as \ac{cpe} through a Wiener process: $\tilde{\phi}[m] \triangleq \frac{1}{K} \sum_{m\bar{K} + L_{\text{c}}}^{m\bar{K} + \bar{K} - 1} \phi[n]$. Since this process is constant over the duration of a symbol, the phase errors are common to all subcarriers. The \ac{ici} coefficient from subcarrier $i$ to $k$ can be expressed as \cite{armstrong1999analysis}
	\begin{align}
		I_{ik} \triangleq \frac{1}{K}\sum_{n=0}^{K-1} \exp\left(j2\pi f_{\text{cfo}} T_{\text{s}} n\right) \exp\left(j2\pi i n/K\right) \exp\left(-j2\pi k n/K\right),
		\label{eq:ici}
	\end{align}
	and packaged in matrix form as $\bm{I}_{\text{ICI}} \in \mathbb{C}^{K \times K}$. The frequency-domain signal for symbol $m$ after \ac{ofdm} processing of (\ref{eq:time_domain_sig}) can then be expressed using (\ref{eq:ici}) as
	\begin{align}
		\tilde{x}^{(m)}[k] = \exp\left(j(2\pi f_{\text{cfo}} T_{\text{sym}} m + \tilde{\phi}[m])\right) \sum_{i=0}^{K-1} I_{ik} {\tilde{h}[i]\tilde{s}^{(m)}[i]} + \tilde{v}^{(m)}[k], \quad k = 0,1,\ldots,K-1.
		\label{eq:freq_domain_sig}
	\end{align}
	When $f_{\text{cfo}} = 0$ and $\tilde{\phi}[m] = 0$, (\ref{eq:freq_domain_sig}) simplifies to
	\begin{align}
		\tilde{x}^{(m)}[k] = \tilde{h}[k]\tilde{s}^{(m)}[k] + \tilde{v}^{(m)}[k].
	\end{align}
	
	Three channel types are considered: (1) an \ac{awgn} channel, for which $L=0$ and
	$h[0]=g$, where $g$ is a deterministic gain; (2) a Rayleigh frequency-flat
	channel, for which $L=0$ and
	$h[0]\sim\mathcal{C}\mathcal{N}(0,\sigma_{h[0]}^2)$; and (3) a Rayleigh
	frequency-selective channel, for which $L\geq1$ and
	$h[l]\sim\mathcal{C}\mathcal{N}(0,\sigma_{h[l]}^2)$ for $l = 0,1,\ldots, L$.
	
	%
	%
	
	\section{Communications Capacity}
	
	As a first step to exploring \ac{ofdm} signal design tradeoffs for ranging and communications, this section explores how the power, quantity, and placement of pilot resources affect \ac{cfo} and \ac{cpe} estimation accuracy, which, in turn, determines the extent to which residual \ac{ici} and phase error present in the received signal limit the communications rate.
	
	\subsection{Phase Errors and Intercarrier Interference}
	
	\ac{ofdm} is particularly susceptible to \ac{cfo} errors, which cause subcarriers to no longer remain orthogonal. This loss of orthogonality causes \ac{ici} and results in a decreased \ac{sinr} in the receiver \cite{stamoulis2002intercarrier}. Additionally, \ac{cpe} arises due to oscillator errors \cite{robertson1995analysis}. Both of these impairments can be estimated and corrected through the use of training sequences and pilots. This subsection derives expressions for the \ac{cfo} and \ac{cpe} estimation errors, the residual \ac{ici} after correction, and the residual \ac{cpe} after correction.
	
	
	Assuming that the receiver experiences a \ac{cfo} of $f_{\text{cfo}}$ \SI{}{\hertz}, and that the Wiener phase noise process $\tilde{\phi}[m]$ has independent Gaussian increments with $\tilde{\phi}[m+1] - \tilde{\phi}[m] \sim \mathcal{N}(0,\sigma_{\phi}^2)$, the received signal model in (\ref{eq:freq_domain_sig}) can be rewritten to separate the desired signal from the interference:
	\begin{align}
		\tilde{x}^{(m)}[k] &= \exp\left(j(2\pi f_{\text{cfo}} T_{\text{sym}} m + \tilde{\phi}[m])\right) \sum_{i=0}^{K-1} I_{ik} {\tilde{h}[i]\tilde{s}^{(m)}[i]} + \tilde{v}^{(m)}[k] \nonumber \\
		&= \exp\left(j(2\pi f_{\text{cfo}} T_{\text{sym}} m + \tilde{\phi}[m])\right) \left( I_{kk}\tilde{h}[k]\tilde{s}^{(m)}[k] + \sum_{i=0, i\neq k}^{K-1} I_{ik} {\tilde{h}[i]\tilde{s}^{(m)}[i]}\right) + \tilde{v}^{(m)}[k] \nonumber \\
		&= \exp\left(j(2\pi f_{\text{cfo}} T_{\text{sym}} m + \tilde{\phi}[m])\right) I_{kk}\tilde{h}[k]\tilde{s}^{(m)}[k] + \tilde{v}_{\text{total}}^{(m)}[k],
		\label{eq:sig_freq_domain}
	\end{align}
	with $\tilde{v}^{(m)}_{\text{total}}[k] \triangleq \tilde{v}^{(m)}_{\text{ICI}}[k]  + \tilde{v}^{(m)}[k]$ and $\tilde{v}^{(m)}_{\text{ICI}}[k] \triangleq \exp(j(2\pi f_{\text{cfo}} T_{\text{sym}} m + \tilde{\phi}[m])) \sum_{i=0, i\neq k}^{K-1} I_{ik} {\tilde{h}[i]\tilde{s}^{(m)}[i]}$.
	
	To simplify expressions, it will be helpful to introduce a few new symbols. Define the average received signal power at subcarrier $k$ during symbol $m$ as $P^{(m)}_k \triangleq \sigma_{\tilde{s}^{(m)}}^2[k] \mathbb{E}[|\tilde{h}[k]|^2]$, and the normalized fading power as $\gamma_{k} \triangleq |\tilde{h}[k]|^2 / \mathbb{E}[|\tilde{h}[k]|^2]$. Define the \ac{ici} power on subcarrier $k$ as $P^{(m)}_{\text{ICI},k} \triangleq \sum_{i=0, i\neq k}^{K-1} |I_{ik}|^2 |\tilde{h}[k]|^2 \sigma_{\tilde{s}^{(m)}}^2[k] = \sum_{i=0, i\neq k}^{K-1} |I_{ik}|^2 P^{(m)}_k \gamma_{k}$.
	Treating the signal on each subcarrier as an independent Gaussian sample, let $\tilde{\bm{s}}^{(m)} \sim \mathcal{C}\mathcal{N}\left(\bm{0},\bm{\Sigma}^{(m)}_s\right)$, where $\bm{\Sigma}^{(m)}_s = \text{diag}\left(\left[\sigma_{\tilde{s}^{(m)}}^2[0],\; \ldots, \sigma_{\tilde{s}^{(m)}}^2[K-1]\right]\right)$. In practice, the symbols $\tilde{\bm{s}}^{(m)}$ will not be Gaussian and will instead adhere to some constellation. However, \ac{ici} can be appropriately modeled as Gaussian by the central limit theorem \cite{russell1995interchannel}, so only the 2nd moments of the symbols $\tilde{\bm{s}}^{(m)}$ will be considered in this analysis. Finally, let $\bm{D}_{\tilde{h}} = \text{diag}([\tilde{h}[0],\; \ldots \tilde{h}[K-1]])$.
	
	With these preliminaries, conditioned on the channel coefficients and \ac{cfo}, the \ac{ici} can be expressed in vector form as $\tilde{\bm{v}}^{(m)}_{\text{ICI}} = \exp\left(j(2\pi f_{\text{cfo}} T_{\text{sym}} m + \tilde{\phi}[m])\right) (\bm{I}_{\text{ICI}} - \bm{I}) \bm{D}_{\tilde{h}} \tilde{\bm{s}}^{(m)}$. Since $\tilde{\bm{s}}^{(m)}$ is Gaussian distributed, it follows that $\bm{\Sigma}_{\text{ICI}}^{(m)} = (\bm{I}_{\text{ICI}} - \bm{I}) \bm{D}_{\tilde{h}} \bm{\Sigma}^{(m)}_s \bm{D}_{\tilde{h}}^H (\bm{I}_{\text{ICI}} - \bm{I})^H$. Therefore, $\tilde{\bm{v}}_{\text{total}}^{(m)} \sim \mathcal{C}\mathcal{N}\left(\bm{0},\bm{\Sigma}^{(m)}_{\tilde{v}_{\text{total}}}\right)$, where $\bm{\Sigma}^{(m)}_{\tilde{v}_{\text{total}}} \triangleq \bm{\Sigma}_{\text{ICI}}^{(m)} + \sigma^2_v\bm{I}$.
	Each element of this total noise is distributed as $\tilde{v}^{(m)}_{\text{total}}[k] \sim \mathcal{C}\mathcal{N}\left(0, \sigma_{\tilde{v}}^2 + P^{(m)}_{\text{ICI},k}\right)$.
	Additionally, the gain caused by attenuation due to \ac{ici} is defined as $P_{\text{att}} = |I_{kk}|^2$, which has no dependence on $k$. The instantaneous \ac{sinr} can then be written as
	\begin{align}
		\text{SINR}^{(m)}_k = \frac{P^{(m)}_k P_{\text{att}} \gamma_k}{P^{(m)}_{\text{ICI},k} + \sigma_{\tilde{v}}^2},
		\label{eq:sinr}
	\end{align}
	which takes a similar form to the \ac{sinr} expressions in \cite{hamdi2010exact,stamoulis2002intercarrier}.
	
	To isolate the \ac{cfo} and \ac{cpe} terms, the effect of the channel coefficients must be mitigated by exploiting the known pilot modulations. Under the block fading assumption, $\tilde{h}[i]$ remains constant for all $m$. Therefore, the receiver can divide the received signal by the known pilot modulations for any resources at symbol $m$ and subcarriers $k \in \mathcal{I}_{\text{pilot}}^{(m)}$. Let the received signal in (\ref{eq:sig_freq_domain}) divided by the known pilot modulation be
	\begin{align}
		\tilde{y}^{(m)}[k] \triangleq \frac{\tilde{x}^{(m)}[k]}{\tilde{s}^{(m)}[k]},
		\label{eq:sig_divided}
	\end{align}
	for $k \in \mathcal{I}_{\text{pilot}}^{(m)}$. Following $\cite{Tretter1985}$, the additive noise $\tilde{v}^{(m)}_{\text{total}}[k]$ can be approximated as phase noise\footnote{In reality, the phase noise will follow the distribution in \cite{luo2020analysis}. The Gaussian approximation is accurate at high SNR, and numerical analysis shows that the approximation's variance is less than the true variance but within \SI{35.5}{\percent} of the true variance for SNRs greater than \SI{-5}{\decibel}. For SNRs lower than \SI{-5}{\decibel}, the approximation becomes increasingly inaccurate. However, communications services are not typically provided at such low SNRs.}
	$\tilde{v}^{(m)}_{\phi}[k]$. In vector form, this is expressed as $\tilde{\bm{v}}^{(m)}_{\phi} \sim \mathcal{N}\left(\bm{0}, \bm{\Sigma}^{(m)}_{\tilde{v}_{\phi}}\right)$, with $\bm{\Sigma}^{(m)}_{\tilde{v}_{\phi}}$ derived in Appendix \ref{app:pn}. The noise vectors from all symbols can be stacked into $\tilde{\bm{v}}_{\phi} \sim \mathcal{N}\left(\bm{0}, \bm{\Sigma}_{\tilde{v}_{\phi}}\right)$, where $\bm{\Sigma}_{\tilde{v}_{\phi}}$ is a block-diagonal matrix consisting of submatrices $\bm{\Sigma}^{(0)}_{\tilde{v}_{\phi}},\; \bm{\Sigma}^{(1)}_{\tilde{v}_{\phi}},\; \ldots,\bm{\Sigma}^{(M-1)}_{\tilde{v}_{\phi}}$. Additionally, the \ac{cpe} process $\phi[m]$ can be written in vector form as $\tilde{\bm{\phi}} \sim \mathcal{C}\mathcal{N}\left(\bm{0},\bm{\Sigma}_{\tilde{\phi}}\right)$, where $\left(\bm{\Sigma}_{\tilde{\phi}}\right)_{nm} = \sigma_{\phi}^2 \min\{n,m\}$.
	
	Some pilot resources will contain a follow-on pilot resource placed at the same subcarrier but in a subsequent symbol. For each $m$ and each $k \in \mathcal{I}_{\text{pilot}}^{(m)}$, let $n_{mk}$ be the nearest subsequent \ac{ofdm} symbol index also containing a pilot resource at subcarrier $k$. Symbolically, $n_{mk} = \inf\{n>m : k \in \left(\mathcal{I}_{\text{pilot}}^{(m)} \cap \mathcal{I}_{\text{pilot}}^{(n)}\right)\}$. For $k \in \mathcal{I}_{\text{pilot}}^{(m)}$ and $n_{mk} < \infty$, a phase difference measurement between a pilot resource and its follow-on pilot resource can be obtained from (\ref{eq:sig_divided}) and expressed as
	\begin{align}
		\Delta\angle \tilde{y}^{(m)}[k] = \angle \tilde{y}^{(n_{mk})}[k] - \angle \tilde{y}^{(m)}[k],
		\label{eq:phase_diff}
	\end{align}
	The number of symbols between each pilot resource and its follow-on pilot resource is defined as $d^{(m)}[k] = n_{mk} - m$. Define $\bm{u}_{mk} \in \mathbb{R}^{M \times 1}$ such that $\bm{u}_{mk}$ is all zero except $u_{mk}[m] = -1$ and $u_{mk}[n_{mk}] = 1$. Then, $\tilde{\phi}[n_{mk}] - \tilde{\phi}[m] = \bm{u}_{mk}^T \tilde{\bm{\phi}}$. Additionally, define $\bm{z}_{mk} \in \mathbb{R}^{MK \times 1}$ such that $\bm{z}_{mk}$ is all zero except $z_{mk}[mK + k] = -1$ and $z_{mk}[n_{mk}K + k] = 1$. Then, $\tilde{v}^{(n_{mk})}_{\phi}[k] - \tilde{v}^{(m)}_{\phi}[k] = \bm{z}_{mk}^T \tilde{\bm{v}}_{\phi}$. As a result, the phase difference measurement in (\ref{eq:phase_diff}) may be expressed as
	\begin{align}
		\Delta\angle \tilde{y}^{(m)}[k] = 2\pi f_{\text{cfo}} T_{\text{sym}} d^{(m)}[k] + \bm{u}_{mk}^T \tilde{\bm{\phi}} + \bm{z}_{mk}^T \tilde{\bm{v}}_{\phi}, \quad k \in \mathcal{I}_{\text{pilot}}^{(m)},\quad n_{mk} < \infty.
		\label{eq:phase_diff_2}
	\end{align}
	The phase difference measurements $\Delta\angle \tilde{y}^{(m)}[k]$ can be stacked as $\Delta\angle \tilde{\bm{y}}^{(m)}$ and the symbol index differences $d^{(m)}[k]$ as $\bm{d}^{(m)}$. Each vector has dimension $N_{\Delta}^{(m)} \times 1$, where $N_{\Delta}^{(m)} = |\{n_{mk} : n_{mk} < \infty\}|$. Stack the row-vectors $\bm{u}_{mk}^T$ to create $\bm{U}^{(m)} \in \mathbb{R}^{N_{\Delta}^{(m)} \times M}$, the measurement matrix for $\tilde{\bm{\phi}}$. Also stack the row-vectors $\bm{z}_{mk}^T$ to create $\bm{Z}^{(m)} \in \mathbb{R}^{N_{\Delta}^{(m)} \times MK}$, the measurement matrix for $\tilde{\bm{v}}_{\phi}$. This results in the vectorized form of (\ref{eq:phase_diff_2}),
	\begin{align}
		\Delta\angle \tilde{\bm{y}}^{(m)} = 2\pi f_{\text{cfo}} T_{\text{sym}} \bm{d}^{(m)} +  \bm{U}^{(m)} \tilde{\bm{\phi}} + \bm{Z}^{(m)} \tilde{\bm{v}}_{\phi}.
		\label{eq:phase_diff_vec}
	\end{align}
	
	Now that the phase differences are expressed in a linear form, \ac{lmmse} estimates \cite{y_barshalom01_tan} of $f_{\text{cfo}}$ and $\tilde{\bm{\phi}}$ can be obtained. Stacking the vector measurements in (\ref{eq:phase_diff_vec}) for all $m$ results in
	\begin{align}
		\begin{bmatrix}
			\Delta\angle \tilde{\bm{y}}^{(0)} \\
			\Delta\angle \tilde{\bm{y}}^{(1)} \\
			\vdots \\
			\Delta\angle \tilde{\bm{y}}^{(M-1)} 
		\end{bmatrix}
		=
		\begin{bmatrix}
			2\pi T_{\text{sym}} \bm{d}^{(0)} \\
			2\pi T_{\text{sym}} \bm{d}^{(1)} \\
			\vdots \\
			2\pi T_{\text{sym}} \bm{d}^{(M-1)} 
		\end{bmatrix}
		f_{\text{cfo}}
		+
		\begin{bmatrix}
			\bm{U}^{(0)} \\
			\bm{U}^{(1)} \\
			\vdots \\
			\bm{U}^{(M-1)} 
		\end{bmatrix} \tilde{\bm{\phi}}
		+
		\begin{bmatrix}
			\bm{Z}^{(0)} \\
			\bm{Z}^{(1)} \\
			\vdots \\
			\bm{Z}^{(M-1)} 
		\end{bmatrix} \tilde{\bm{v}}_{\phi},
		\label{eq:cfo_est}
	\end{align}
	which may be written compactly as $\Delta\angle\tilde{\bm{y}} = \bm{t}f_{\text{cfo}} + \bm{U} \tilde{\bm{\phi}} + \bm{Z} \tilde{\bm{v}}_{\phi}$.
	
	Define the stacked parameter vector $\bm{\beta} \triangleq [f_{\text{cfo}},\tilde{\bm{\phi}}^T]^T$. Recall that the prior covariance of $\tilde{\bm{\phi}}$ is $\bm{\Sigma}_{\tilde{\phi}}$. Since the receiver has no prior for $f_{\text{cfo}}$, a diffuse prior is assumed such that $f_{\text{cfo}} \sim \mathcal{N}(0,\sigma_{\text{cfo}}^2)$ and $\sigma_{\text{cfo}} \to \infty$. Define $\bm{\Sigma}_{\text{weight}}$ as the limit of the inverse of the prior covariance of $\bm{\beta}$,
	\begin{align}
		\bm{\Sigma}_{\text{weight}} \triangleq \lim_{\sigma_{\text{cfo}}\to\infty} \bm{\Sigma}_{\beta}^{-1} =
		\begin{bmatrix}
			0 & \ldots \\
			\vdots & \bm{\Sigma}_{\tilde{\phi}}^{-1}
		\end{bmatrix}.
	\end{align}
	Defining $\bm{A} \triangleq [\bm{t},\bm{U}]$ and $\bm{\Sigma}_{\Delta \tilde{v}_{\phi}} \triangleq \bm{Z}\bm{\Sigma}_{\tilde{v}_{\phi}}\bm{Z}^T$, the \ac{lmmse} estimate of $\bm{\beta}$ takes the form
	\begin{align}
		\hat{\bm{\beta}} = \left( \bm{\Sigma}_{\text{weight}} + \bm{A}^{T} \bm{\Sigma}_{\Delta \tilde{v}_{\phi}}^{-1}\bm{A}\right)^{-1}\bm{A}^{T} \bm{\Sigma}_{\Delta \tilde{v}_{\phi}}^{-1} (\Delta\angle\tilde{\bm{y}}),
	\end{align}
	resulting in an estimation error $\bm{\epsilon}_{\beta} = \hat{\bm{\beta}} - \bm{\beta}$ with covariance $\bm{\Sigma}_{\epsilon_{\beta}} = \left( \bm{\Sigma}_{\text{weight}} + \bm{A}^{T} \bm{\Sigma}_{\Delta \tilde{v}_{\phi}}^{-1} \bm{A}\right)^{-1}$. Estimates $\hat{f}_{\text{cfo}}$ of $f_{\text{cfo}}$ and $\hat{\tilde{\bm{\phi}}}$ of $\tilde{\bm{\phi}}$ are extracted from $\hat{\bm{\beta}}$, yielding estimation errors $\epsilon_\text{cfo} = \hat{f}_{\text{cfo}} - f_{\text{cfo}}$ and $\bm{\epsilon}_{\tilde{\phi}} = \hat{\tilde{\bm{\phi}}} - \tilde{\bm{\phi}}$. \ac{cfo} estimation error normalized by the subcarrier spacing is defined as $\delta \triangleq K T_{\text{s}} \epsilon_{\text{cfo}}$ and will be used in the \ac{ici} analysis. Element-wise, the \ac{cpe} estimation errors are denoted $\epsilon_{\tilde{\phi}}[m] = \hat{\tilde{\phi}}[m] - \tilde{\phi}[m]$.
	
	Having obtained \ac{cfo} and \ac{cpe} estimates, the receiver can correct for the estimated \ac{cfo} in its time-domain signal in (\ref{eq:time_domain_sig}) by multiplying $x[n]$ by $\exp\left(-j2\pi \hat{f}_{\text{cfo}} T_{\text{s}} n \right)$, perform DFT processing, and then correct for the estimated \ac{cpe} in its frequency-domain signal by multiplying the $m$th symbol by $\exp\left(-j\hat{\tilde{\phi}}[m]\right)$. Let $\epsilon_{\tilde{\phi},\text{total}}[m] \triangleq 2\pi \delta T_{\text{sym}} m / (K T_{\text{s}}) + \epsilon_{\tilde{\phi}}[m]$. This correction results in updated \ac{ici} coefficients $I_{ik} = \frac{1}{K}\sum_{n=0}^{K-1} \exp\left(j(2\pi \delta n/K)\right) \exp\left(j2\pi i n/K\right)  \exp\left(-j2\pi k n/K\right)$ and updated total interference-and-noise vectors
	\begin{align}
		\tilde{v}^{(m)}_{\text{total}}[k] = \exp\left(j\epsilon_{\tilde{\phi},\text{total}}[m]\right) \left(\sum_{i=0, i\neq k}^{K-1} I_{ik} {\tilde{h}[i]\tilde{s}^{(m)}[i]}\right) + \tilde{v}_{\text{corr}}^{(m)}[k],
	\end{align}
	where $\tilde{v}^{(m)}_{\text{corr}}[k] \sim\mathcal{C}\mathcal{N}(0,\sigma_{v}^2)$ is the frequency-domain noise after correction, distributed with the same variance $\sigma_{v}^2$ as the noise in (\ref{eq:time_domain_sig}). Using these updated quantities, the corrected signal after the DFT is
	\begin{align}
		\tilde{x}^{(m)}_{\text{corr}}[k] &= \exp\left(j\epsilon_{\tilde{\phi},\text{total}}[m]\right) I_{kk}\tilde{h}[k]\tilde{s}^{(m)}[k] + \tilde{v}^{(m)}_{\text{total}}[k].
		\label{eq:final_corr_sig}
	\end{align}
	
	Now that the receiver has corrected for \ac{cfo} and \ac{cpe}, it can finally estimate the channel. This channel estimation step is also prone to errors which will depend upon the power, quantity, and placement of the pilots.

	\subsection{Channel Estimation and Capacity}
	\label{sec:capacity}
	This paper's approach to computing the channel capacity is inspired by \cite{Ohno2004}, which considers channel estimation error present in the \ac{lmmse} estimate of the channel coefficients $h[l]$. This estimation error factors into the communication link's effective \ac{sinr}.
	
	\subsubsection{Estimating Channel Coefficients}
	
	For simplicity of expression, the residual symbol-dependent phase errors will be dropped from the signal model in (\ref{eq:final_corr_sig}) for analyzing channel estimation error\footnote{In practice, residual symbol-dependent phase errors will remain present during channel estimation. While large phase errors could cause symbols to become incoherent and increase channel estimation errors, a phase rotation limit is defined and enforced later in Sec.~\ref{sec:channel_capacity} such that capacity is only computed when residual phase errors are small. Dropping the residual phase errors still results in optimistic channel estimation errors but is an appropriate approximation since these phase errors are small.}, resulting in the simplified model
	\begin{align}
		\tilde{x}^{(m)}_{\text{corr}}[k] &\approx I_{kk}\tilde{h}[k]\tilde{s}^{(m)}[k] + \tilde{v}^{(m)}_{\text{total}}[k] \nonumber \\
		&= \sqrt{P_{\text{att}}}\tilde{h}_{\text{corr}}[k]\tilde{s}^{(m)}[k] + \tilde{v}^{(m)}_{\text{total}}[k],
		\label{eq:simple_sig}
	\end{align}
	where $P_{\text{att}} = |I_{kk}|^2$, and $\tilde{h}_{\text{corr}}[k] \triangleq \exp(j\angle I_{kk}) \tilde{h}[k]$ and is distributed identically to $\tilde{h}[k]$. This signal can be written in vector form as $\tilde{\bm{x}}_{\text{corr}} ^{(m)} = \sqrt{P_{\text{att}}} \tilde{\bm{h}}_{\text{corr}} \odot \tilde{\bm{s}}^{(m)} + \tilde{\bm{v}}^{(m)}_{\text{total}}$.
	
	Assume that the receiver knows $L$ and has statistical knowledge of the channel coefficient distributions. Following \cite{Ohno2004}, a \ac{lmmse} channel estimation error variance at subcarrier $k$ is determined and denoted $\sigma_{\epsilon_{\tilde{h}}}^2[k]$. Define the variance of the frequency-domain channel coefficients as $\sigma_{\tilde{h}}^2 = \mathbb{E}[|\tilde{h}_{\text{corr}}[k]|^2]$. Treating the channel estimation errors as additive noise, an effective \ac{sinr} at subcarrier $k$ and symbol $m$ is defined as
	\begin{align}
		\rho_k^{(m)} &= \frac{\left(\sigma_{\tilde{h}}^2 - \sigma_{\epsilon_{\tilde{h}}}^2[k]\right) \gamma_{k}}{\sigma_{\epsilon_{\tilde{h}}}^2[k] + \frac{\sigma_{\tilde{v}}^2 + P^{(m)}_{\text{ICI},k}}{\sigma_{\tilde{s}^{(m)}}^2[k] P_{\text{att}}}}.
		\label{eq:rho}
	\end{align}
	When $\sigma_{\epsilon_{\tilde{h}}}^2[k] = 0$, (\ref{eq:rho}) simplifies to
	\begin{align}
		\rho_k^{(m)} = \frac{P^{(m)}_k P_{\text{att}} \gamma_{k}}{P^{(m)}_{\text{ICI},k} + \sigma_{\tilde{v}}^2},
		\label{eq:rho2}
	\end{align}
	which is identical to the expression for \ac{sinr} in (\ref{eq:sinr}). The effective \ac{sinr} in (\ref{eq:rho}) factors in all of the modeled impairments and will be used when computing channel capacity.
	
	\subsubsection{Maximum Ratio Combining}
	
	The signal modeling and analysis up to this point have only considered single-antenna receivers. Now consider a receiver with $N_{\text{RX}}$ receive antennas. Assume that the channel coefficients for each antenna are i.i.d. and that each antenna's signal experiences the same \ac{cfo} and \ac{cpe}. The channel coefficients are denoted $\bm{h}_i$ for $i \in [0, \ldots, N_{\text{RX}}-1]$. Under these assumptions, the receiver's \ac{cfo} and \ac{cpe} estimates can be improved by stacking the phase difference measurements in (\ref{eq:cfo_est}) from each antenna. After estimating and correcting for the \ac{cfo} and \ac{cpe}, the channels are independently estimated for each antenna. After channel estimation, the receiver can employ \ac{mrc} \cite{GoldsmithBook}, improving the effective \ac{sinr}. Denote the individual effective \ac{sinr} for antenna $i$ at subcarrier $k$ and symbol $m$ as $\rho_{ki}^{(m)}$ for $i = 0,1,\ldots,N_{\text{RX}}-1$, using (\ref{eq:rho}). Then the \ac{mrc} \ac{sinr} is $\rho_{\text{MRC},k}^{(m)} = \sum_{i=0}^{N_{\text{RX}}-1} \rho_{ki}^{(m)}$.
	
	\subsubsection{Computing Channel Capacity}
	\label{sec:channel_capacity}
	Now that an expression has been derived for the effective \ac{sinr} that accounts for \ac{cfo}, \ac{cpe}, and channel estimation errors, channel capacity and outage probability can be quantified. Given $\rho_{k}^{(m)}$, the \ac{sinr} on subcarrier $k$ during symbol $m$, the instantaneous capacity \cite{GoldsmithBook} summed over all resources and normalized by the number of samples \cite{Ohno2004} is
	\begin{align}
		C_0 \triangleq \frac{1}{M\bar{K}} \sum_{m=0}^{M-1} \sum_{k \in \mathcal{I}_{\text{data}}^{(m)}}{ \log_2\left(1+\rho_{k}^{(m)}\right)}.
		\label{eq:rate_simple}
	\end{align}
	
	However, from (\ref{eq:final_corr_sig}), a residual, symbol-dependent phase rotation remains present in the signal. The effect of uncompensated phase errors on capacity is difficult to analyze without simulating symbol constellations, since its impact on symbol decoding error is fundamentally different from that of \ac{awgn}. While some work has studied this capacity from an information theoretic view \cite{khanzadi2015capacity,lapidoth2002phase}, this paper opts for a simpler treatment. Residual CPE is tolerable so long as the phase rotation is small enough that symbols do not get rotated into incorrect decoding regions. If the phase rotation is large enough, however, the symbol error rate may increase rapidly. To remain agnostic to constellations and account for this behavior, a tolerable limit of phase rotation $\epsilon_{\phi,\text{max}}$ is assumed such that if $|\epsilon_{\tilde{\phi},\text{total}}[m]| > \epsilon_{\phi,\text{max}}$, symbol $m$ is unused and no data is communicated. Define the indicator function of this condition as $\mathbbm{1}_{\epsilon_{\tilde{\phi},\text{total}}}[m]$. Then multiplying (\ref{eq:rate_simple}) by $\mathbbm{1}_{\epsilon_{\tilde{\phi},\text{total}}}[m]$ results in a new expression for the instantaneous capacity
	\begin{align}
		C \triangleq \frac{1}{M\bar{K}} \sum_{m=0}^{M-1} \sum_{k \in \mathcal{I}_{\text{data}}^{(m)}}{ \mathbbm{1}_{\epsilon_{\tilde{\phi},\text{total}}}[m] \log_2\left(1+\rho_{k}^{(m)}\right)}.
		\label{eq:rate}
	\end{align}
	The ergodic channel capacity can then be defined as
	\begin{align}
		\bar{C} \triangleq \mathbb{E}_{\bm{h},\delta,\epsilon_{\tilde{\phi}}[m]}\left[ C \right],
		\label{eq:cap}
	\end{align}
	where the expectation is taken over the channel coefficients $\bm{h}$, the residual \ac{cfo} $\delta$, and the residual phase error $\epsilon_{\tilde{\phi}}[m]$. In addition to capacity, the probability of outage is quantified, i.e., the probability that the instantaneous capacity drops below a particular rate requirement. Given a rate requirement $C_{\text{min}}$, this is
	\begin{align}
		P_\text{outage} \triangleq P\left(C < C_{\text{min}}\right).
		\label{eq:out}
	\end{align}
	If $N_{\text{RX}} > 1$, the receiver uses \ac{mrc}. Then the instantaneous capacity is
	\begin{align}
		C_{\text{MRC}} \triangleq \frac{1}{M\bar{K}} \sum_{m=0}^{M-1} \sum_{k \in \mathcal{I}_{\text{data}}^{(m)}}{ \mathbbm{1}_{\epsilon_{\tilde{\phi},\text{total}}}[m] \log_2\left(1+\rho_{\text{MRC},k}^{(m)}\right)},
		\label{eq:rate_mrc}
	\end{align}
	the ergodic capacity is
	\begin{align}
		\bar{C}_{\text{MRC}} \triangleq \mathbb{E}_{\bm{h}_0,\; \ldots,\bm{h}_{N_\text{RX}-1} ,\delta,\epsilon_{\tilde{\phi}}[m]}\left[ C_{\text{MRC}} \right],
		\label{eq:cap_mrc}
	\end{align}
	and the probability of outage is
	\begin{align}
		P_\text{outage,MRC} \triangleq P\left(C_{\text{MRC}} < C_{\text{min}}\right).
		\label{eq:out_mrc}
	\end{align}
	
	Equations (\ref{eq:rate}) through (\ref{eq:out_mrc}) are key to understanding the channel capacity and outage probability as a function of arbitrarily-placed pilot resources in \ac{ofdm}-based communications. Such expressions, accounting fully for the signal-dependent estimation errors of \ac{cfo}, \ac{cpe}, and channel coefficients, are novel to the best of the authors' knowledge.
	
	%
	%
	
	\section{Ranging Variance}
	
	The next step in quantifying the \ac{ofdm} signal design tradeoffs for ranging and communications is to consider the factors that affect the receiver's range estimation variance. Specifically, this section explores how the power, quantity, and placement of pilot resources impact the Ziv-Zakai bound on \ac{toa} estimate variance.
	
	\begin{figure}[h!]
		\centering
		\includegraphics[width=0.5\textwidth]{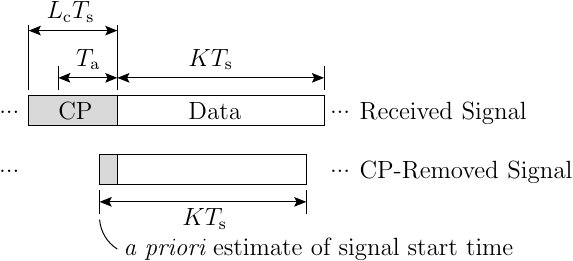}
		\caption{A graphical depiction of the received signal and the signal after cyclic prefix removal. The receiver has \textit{a priori} knowledge of the starting time of the signal within a duration $T_{\text{a}}$ shorter than the cyclic prefix.}
		\label{fig:toa_prior}
	\end{figure}
	
	The Ziv-Zakai bound will be used to bound the system's ranging variance by providing a lower bound on the variance of the receiver's \ac{toa} estimate. Denote the true \ac{toa} as $\tau$ in units of seconds, the \ac{toa} estimation error as $\epsilon_\tau$, and the Ziv-Zakai bound as ZZB. Then $\mathbb{E}[\epsilon_\tau^2] \geq \text{ZZB}$. To construct the ZZB, consider a binary detection problem where two hypotheses are equally likely: (1) the received signal experienced delay $\tau$, and (2) the received signal experienced delay $\tau+\tau_z$, where $\tau_z = T_{\text{s}} z$, and $z \in \mathbb{R}$ is the additional delay in units of samples. The probability of error of the optimal detector between these two hypotheses is defined as $P_{\text{min}}(z)$. If the receiver has \textit{a priori} knowledge that the \ac{toa} is uniformly distributed in $[0,T_{\text{a}}]$, then the Ziv-Zakai bound can be written as \cite{Ziv1969,Dardari2009}
	\begin{align}
		\text{ZZB} = \frac{1}{T_{\text{a}}} \int_0^{T_{\text{a}}}{\tau_z(T_{\text{a}}-\tau_z) P_{\text{min}}(z) d\tau_z}.
		\label{eq:zzb}
	\end{align}
	
	Derivations of $P_{\text{min}}(z)$ will now be explored for different types of fading and channel knowledge: (1) \ac{awgn}, (2) Rayleigh fading with perfect channel knowledge, (3) Rayleigh fading with perfect channel knowledge and \ac{mrc}, and (4) Rayleigh fading with no channel knowledge. Throughout this section, it is assumed that the receiver's \textit{a priori} \ac{toa} distribution falls within the duration of the cyclic prefix as visualized in Fig.~\ref{fig:toa_prior}, allowing the receiver to discard the cyclic prefix and perform circular correlation against each \ac{ofdm} symbol as in \cite{berger2010signal}.
	
	\subsection{Perfect Channel Knowledge}
	
	Under the first three channel types, the receiver has perfect knowledge of the channel coefficients, the \ac{cfo}, and the \ac{cpe}. As a result, no estimation errors are present, lower-bounding the ranging variance for each channel type.
	
	Define the pilot vector $\tilde{\bm{b}}^{(m)} \in \mathbb{C}^{K}$ such that $\tilde{b}^{(m)}[k] = 0$ for $k \in \mathcal{I}_{\text{data}}^{(m)}$ and $\tilde{b}^{(m)}[k] = \tilde{s}^{(m)}[k]$ for $k \in \mathcal{I}_{\text{pilot}}^{(m)}$. The inverse DFT of $\tilde{\bm{b}}^{(m)}$ is $\bm{b}^{(m)}=\bm{F}^{H}\tilde{\bm{b}}^{(m)}$.
	First consider the \ac{awgn} channel, where the received pilot signal is $\bm{x}_{\text{p}}^{(m)} = \bm{b}^{(m)} + \bm{v}^{(m)}$. Consider a real-valued representation of the signal where the real and imaginary coefficients are stacked:
	$\bm{x}^{(m)}_{\text{sep}} = \bm{\mu}^{(m)} + \bm{\eta}^{(m)}$,
	$\bm{\mu}^{(m)} = [\Re({\bm{b}^{(m)}}^{T}),\; \Im({\bm{b}^{(m)}}^{T})]^{T}$, and
	$\bm{\eta}^{(m)} = [\Re({\bm{v}^{(m)}}^{T}),\; \Im({\bm{v}^{(m)}}^{T})]^{T}$. The variance of each of the $2K$ elements of $\bm{\eta}$ is $\sigma_\eta^2 = \frac{\sigma_v^2}{2}$. Define a circular shift function $f_z(\bm{x})$, which circularly shifts the vector $\bm{x}$ by $z$ samples. Note that $z$ may be a fractional number of samples, in which case sinc interpolation is applied. Without loss of generality, assume $\tau=0$. The \ac{pdf} for the received signal with delay $\tau$ is
	\begin{align}
		p(\bm{x}^{(m)}_{\text{sep}}|\tau) = \frac{\exp(-\frac{1}{2\sigma_{\eta}^2}(\bm{x}^{(m)}_{\text{sep}}-\bm{\mu}^{(m)})^{T} (\bm{x}^{(m)}_{\text{sep}}-\bm{\mu}^{(m)}))}{(2\pi\sigma_{\eta}^2)^{K}},
	\end{align}
	and the \ac{pdf} for the received signal with delay $\tau + \tau_z$ is
	\begin{align}
		p(\bm{x}^{(m)}_{\text{sep}}|\tau+\tau_z) = \frac{\exp(-\frac{1}{2\sigma_{\eta}^2}(\bm{x}^{(m)}_{\text{sep}}-f_z(\bm{\mu}^{(m)}))^{T} (\bm{x}^{(m)}_{\text{sep}}-f_z(\bm{\mu}^{(m)})))}{(2\pi\sigma_{\eta}^2)^{K}}.
	\end{align}
	The likelihood ratio between the delay $\tau$ and delay $\tau + \tau_z$ signals is then defined as
	\begin{align}
		\Lambda \triangleq \frac{\prod_{m=0}^{M-1} p(\bm{x}^{(m)}_{\text{sep}}|\tau)}{\prod_{m=0}^{M-1} p(\bm{x}^{(m)}_{\text{sep}}|\tau+\tau_z)},
	\end{align}
	and the detection test is $\log{}\Lambda > 0$. The detection statistic $\log{}\Lambda$ can be simplified as follows:
	%
	\begin{align}
		\log{}\Lambda &= \sum_{m=0}^{M-1} -\frac{1}{2\sigma_{\eta}^2}(\bm{x}^{(m)}_{\text{sep}}-\bm{\mu}^{(m)})^{T} (\bm{x}^{(m)}_{\text{sep}}-\bm{\mu}^{(m)}) + \frac{1}{2\sigma_\eta^2}(\bm{x}^{(m)}_{\text{sep}}-f_z(\bm{\mu}^{(m)}))^{T} (\bm{x}^{(m)}_{\text{sep}}-f_z(\bm{\mu}^{(m)})) \nonumber \\
		&= \sum_{m=0}^{M-1} \frac{1}{2\sigma_\eta^2} {\bm{x}^{(m)}_{\text{sep}}}^{T}(\bm{\mu}^{(m)}-f_z(\bm{\mu}^{(m)})).
	\end{align}
	
	Without loss of generality, assuming the first hypothesis is true and $\bm{x}^{(m)}_{\text{sep}} = \bm{\mu}^{(m)} + \bm{\eta}^{(m)}$, the test can be rewritten as
	\begin{align}
		\sum_{m=0}^{M-1} {\bm{\eta}^{(m)}}^{T}(\bm{\mu}^{(m)}-f_z(\bm{\mu}^{(m)})) > \sum_{m=0}^{M-1} {\bm{\mu}^{(m)}}^{T}\bm{\mu}^{(m)} - {\bm{\mu}^{(m)}}^{T}f_z(\bm{\mu}^{(m)}).
		\label{eq:simplified_test}
	\end{align}
	
	Denote the real component of the circular autocorrelation function at delay $l$ samples as $a(l) = \sum_{m=0}^{M-1} {\bm{\mu}^{(m)}}^{T}f_l(\bm{\mu}^{(m)})$. Recognize in (\ref{eq:simplified_test}) that $\sum_{m=0}^{M-1} {\bm{\mu}^{(m)}}^{T}\bm{\mu}^{(m)} = a(0)$ is the real component of the circular autocorrelation of $\bm{b}^{(m)}$ evaluated at a delay of $0$ samples, and $\sum_{m=0}^{M-1} {\bm{\mu}^{(m)}}^{T}f_z(\bm{\mu}^{(m)}) = a(z)$ is the real component of the same autocorrelation evaluated at a delay of $z$ samples. Also recognize in (\ref{eq:simplified_test}) that $\sum_{m=0}^{M-1} {\bm{\eta}^{(m)}}^{T}(\bm{\mu}^{(m)}-f_z(\bm{\mu}^{(m)}))$ is a linear combination of jointly-distributed Gaussian random variables and thus is itself distributed as a zero-mean Gaussian with variance
	\begin{align}
		\sigma_{\eta}^2 \sum_{m=0}^{M-1} ||\bm{\mu}^{(m)}-f_z(\bm{\mu}^{(m)})||^2.
	\end{align}
	
	Therefore, the probability of error for the detection test can be rewritten as
	\begin{align}
		P_{\text{min}}(z) = Q\left(\frac{a(0)-a(z)}{\sigma_{\eta}\sqrt{\sum_{m=0}^{M-1} ||\bm{\mu}^{(m)}-f_z(\bm{\mu}^{(m)})||^2}}\right).
	\end{align}
	
	Under Rayleigh fading with perfect channel knowledge, the receiver can distort its known reference signal to account for the channel prior to correlation. The distorted pilot signal can be described as $\check{\bm{b}}^{(m)} = \bm{b}^{(m)} \circledast \bm{h}$, where $\circledast$ is circular convolution. Stack the real and imaginary coefficients as before, yielding $\bm{x}_{\text{sep}}^{(m)} = \check{\bm{\mu}}^{(m)} + \bm{\eta}^{(m)}$ and
	$\check{\bm{\mu}}^{(m)} = [\Re({{}\check{\bm{b}}^{(m)}}^{T}),\; \Im({{}\check{\bm{b}}^{(m)}}^{T})]^{T}$. Let $\check{a}(l) = \sum_{m=0}^{M-1} {{}\check{\bm{\mu}}^{(m)}}^{T}f_l(\check{\bm{\mu}}^{(m)})$ be the real component of the circular correlation function at a delay of $l$ samples. Conditioning on $\bm{h}$, an identical derivation as in the \ac{awgn} case results in
	\begin{align}
		P_{\text{min}}(z|\bm{h}) &= Q\left(\frac{\check{a}(0)-\check{a}(z)}{\sigma_{\eta}\sqrt{\sum_{m=0}^{M-1}\||\check{\bm{\mu}}^{(m)}-f_z(\check{\bm{\mu}}^{(m)})||^2}}\right), \nonumber \\
		P_{\text{min}}(z) &= \mathbb{E}_{\bm{h}}\left[P_{\text{min}}(z|\bm{h})\right].
		\label{eq:pmin_rayleigh_cdf}
	\end{align}
	
	After the expectation over channel realizations in (\ref{eq:pmin_rayleigh_cdf}) is taken, the minimum probability of error can be substituted into (\ref{eq:zzb}) to obtain the Ziv-Zakai bound.
	
	Next, the probability of error will be quantified for the case of Rayleigh fading with perfect channel knowledge and \ac{mrc}. Recall that this case assumes that the receiver has $N_{\text{RX}}$ receive antennas which experience independent multipath Rayleigh channels. The receiver distorts the pilot signal by each antenna's known channel and then performs correlation with a coherent combination across antennas. Let $\bm{h}_i$ for $i \in [0, \ldots, N_{\text{RX}}-1]$ be the channel coefficients at each antenna and $\check{\bm{b}}^{(m)}_i = \bm{b}^{(m)} \circledast \bm{h}_i$ be the distorted pilot signals at receiver $i$. Accordingly, the autocorrelation of the distorted pilot signal for receiver $i$ at delay $l$ is $\check{a}_i(l)$.
	
	\begin{sloppypar}The received signal at receiver $i$ follows as $\bm{x}^{(m)}_{\text{sep},i} = \check{\bm{\mu}}^{(m)}_i + \bm{\eta}^{(m)}_i$ where
	$\check{\bm{\mu}}^{(m)}_i = [\Re({{}\check{\bm{b}}^{(m)}_i}^{T}),\;  \Im({{}\check{\bm{b}}^{(m)}_i}^{T})]^{T}$.
	Since the noise $\bm{\eta}^{(m)}_i$ and channels $\bm{h}_i$ are independent between receive antennas, the same derivation can be followed by conditioning on all channels $\bm{h}_i$, resulting in\end{sloppypar}
	\begin{align}
		P_{\text{min}}(z|\bm{h}_0,\;\ldots, \bm{h}_{N_{\text{RX}}-1}) &= Q\left(\frac{\sum_{i=0}^{N_{\text{RX}}-1} \check{a}_i(0)-\check{a}_i(z)}{\sigma_{\eta}\sqrt{\sum_{i=0}^{N_{\text{RX}}-1} \sum_{m=0}^{M-1} ||\check{\bm{\mu}}_{i}^{(m)}-f_z(\check{\bm{\mu}}_{i}^{(m)})||^2}}\right). \nonumber \\
		P_{\text{min}}(z) &= \mathbb{E}_{\bm{h}_0,\;\ldots, \bm{h}_{N_{\text{RX}}-1}}\left[P_{\text{min}}(z|\bm{h}_0,\;\ldots, \bm{h}_{N_{\text{RX}}-1})\right].
		\label{eq:pmin_rayleigh_mrc_cdf}
	\end{align}
	%
	%
	
	\subsection{No Channel Knowledge}
	
	\fussy So far, the derivations for the Ziv-Zakai bound have assumed that the receiver performing ranging has perfect channel knowledge. While important for understanding the fundamental limits of ranging precision, such information may not be available to the receiver. A new expression for $P_{\text{min}}(z)$ will be derived where the receiver has no channel knowledge, but has statistical knowledge of the fading distribution. Without explicit channel knowledge, a practical receiver would also be susceptible to the errors caused by \ac{cfo} and \ac{cpe}. However, these errors will not be treated in this expression for  $P_{\text{min}}(z)$, simplifying the analysis and yielding a lower-bound that only captures the effects of unknown channel coefficients.
	
	Let $\bm{B}^{(m)} \in \mathbb{C}^{K\times{}\bar{L}}$ be a Toeplitz matrix whose $l$th column is equal to $\bm{b}^{(m)}$ circularly shifted by $l$ samples for $l = 0,1,\ldots, L$. Following \cite{Ohno2004}, the received pilot signal can be expressed as $\bm{x}_{\text{p}}^{(m)}= \bm{B}^{(m)}\bm{h}+\bm{v}^{(m)}$. Additionally, let $f_z(\bm{B})$ indicate a circular shift of the columns of $\bm{B}$ by $z$ samples. Without any channel corrections applied, the received signal is zero-mean and has a covariance $\bm{\Sigma}_{x}$, which is a block-matrix whose $n$th row partition and $m$th column partition are
	\begin{align}
		\bm{\Sigma}_{x}^{(nm)} = \bm{B}^{(n)}\bm{\Sigma}_h{\bm{B}^{(m)}}^{H} + \mathbbm{1}_{\{n=m\}}\bm{\Sigma}_v.
	\end{align}

	The real and imaginary coefficients are stacked, creating $\bm{x}_{\text{sep}} = [\mathfrak{R}(\bm{x}_{\text{p}}^T),\; \mathfrak{I}(\bm{x}_{\text{p}}^T)]^T$ with covariance $\bm{\Sigma}_{x_{\text{sep}}} = \left[\begin{smallmatrix}\mathfrak{R}(\bm{\Sigma}_{x}) & -\mathfrak{I}(\bm{\Sigma}_{x})\\ \mathfrak{I}(\bm{\Sigma}_{x}) & \mathfrak{R}(\bm{\Sigma}_{x})\end{smallmatrix}\right]$. The \ac{pdf} for $\bm{x}_{\text{sep}}$ with delay $\tau$ is
	\begin{align}
		p(\bm{x}_{\text{sep}}|\tau) = \prod_{m=0}^{M-1} \frac{\exp(-\frac{1}{2}\bm{x}_{\text{sep}}^{T} \bm{\Sigma}_{x_{\text{sep}}}^{-1} \bm{x}_{\text{sep}})}{(2\pi)^{K}|\bm{\Sigma}_{x_{\text{sep}}}|^{\frac{1}{2}}},
	\end{align}
	and the \ac{pdf} with delay $\tau+\tau_z$ is
	\begin{align}
		p(\bm{x}_{\text{sep}}|\tau+\tau_z) = \prod_{m=0}^{M-1} \frac{\exp(-\frac{1}{2}\bm{x}_{\text{sep}}^{T} \bm{\Sigma}_{x_{\text{sep},z}}^{-1} \bm{x}_{\text{sep}})}{(2\pi)^{K}|\bm{\Sigma}_{x_{\text{sep},z}}|^{\frac{1}{2}}},
	\end{align}
	which uses a new covariance $\bm{\Sigma}_{x_{\text{sep},z}} = \left[\begin{smallmatrix}\mathfrak{R}(\bm{\Sigma}_{x_z}) & -\mathfrak{I}(\bm{\Sigma}_{x_z})\\ \mathfrak{I}(\bm{\Sigma}_{x_z}) & \mathfrak{R}(\bm{\Sigma}_{x_z})\end{smallmatrix}\right]$, where $\bm{\Sigma}_{x_z}$ is a block-matrix whose $n$th row partition and $m$th column partition are
	\begin{align}
		\bm{\Sigma}^{(nm)}_{x_z} =  f_z(\bm{B}^{(n)})\bm{\Sigma}_h{f_z(\bm{B}^{(m)})}^{H} + \mathbbm{1}_{\{n=m\}}\bm{\Sigma}_v.
	\end{align}
	%
	The optimal detection test between these two hypotheses takes the form
	\begin{align}
		\log \Lambda = \bm{x}_{\text{sep}}^{T}\bm{Q}_z\bm{x}_{\text{sep}} + \frac{1}{2}\log\frac{|\bm{\Sigma}_{x_{\text{sep},z}}|}{|\bm{\Sigma}_{x_{\text{sep}}}|} > 0, \quad \bm{Q}_z = \frac{1}{2}\left(\bm{\Sigma}_{x_{\text{sep},z}}^{-1} - \bm{\Sigma}_{x_{\text{sep}}}^{-1}\right).
	\end{align}
	This form matches that of \cite[eq. 3]{das2021method}. The minimum probability of error can then be defined as
	\begin{align}
		P_{\text{min}}(z) = P\left({\bm{x}_{\text{sep}}}^{T}\bm{Q}_z\bm{x}_{\text{sep}} + \frac{1}{2}\log\frac{|\bm{\Sigma}_{x_{\text{sep},z}}|}{|\bm{\Sigma}_{x_{\text{sep}}}|} < 0\right).
		\label{eq:pmin_cdf_none}
	\end{align}
	This quadratic form has a generalized chi-squared distribution, which has no closed-form expression for its \ac{cdf}. However, this \ac{cdf} can be computed using Imhof's method \cite{imhof1961computing} or Monte-Carlo methods.
	
	\section{Results}
	\label{sec:results}
	
	Simulated analysis of two example scenarios is now provided to illustrate the use of the ergodic capacity bounds in (\ref{eq:cap}) and (\ref{eq:cap_mrc}), the outage probabilities in (\ref{eq:out}) and (\ref{eq:out_mrc}), and the Ziv-Zakai ranging variance bound in (\ref{eq:zzb}). These results highlight the tradeoffs between communication capacity and ranging variance as a function of pilot placements, pilot power allocations, and propagation environments.
	
	\subsection{Simulation Setup}
	
\begin{figure*}[h!]
		\centering
		\begin{subfigure}{0.48\linewidth}
			\centering
			\includegraphics[width=\textwidth,page=7]{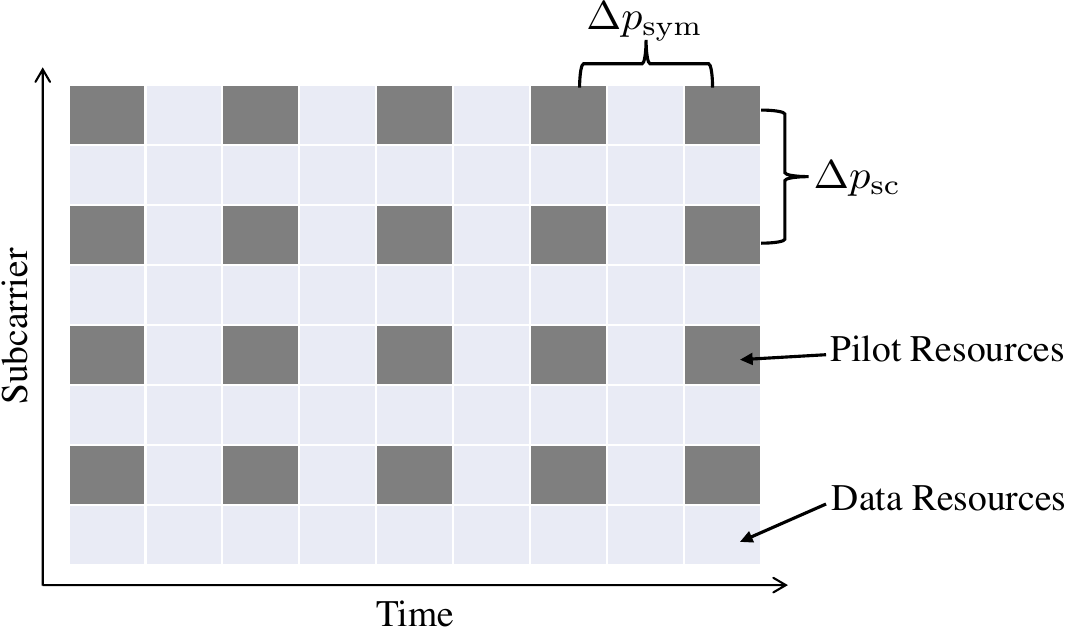}
			\caption{}
			\label{fig:signal_1}
		\end{subfigure}
		\begin{subfigure}{0.48\linewidth}
			\centering
			\includegraphics[width=\textwidth,page=8]{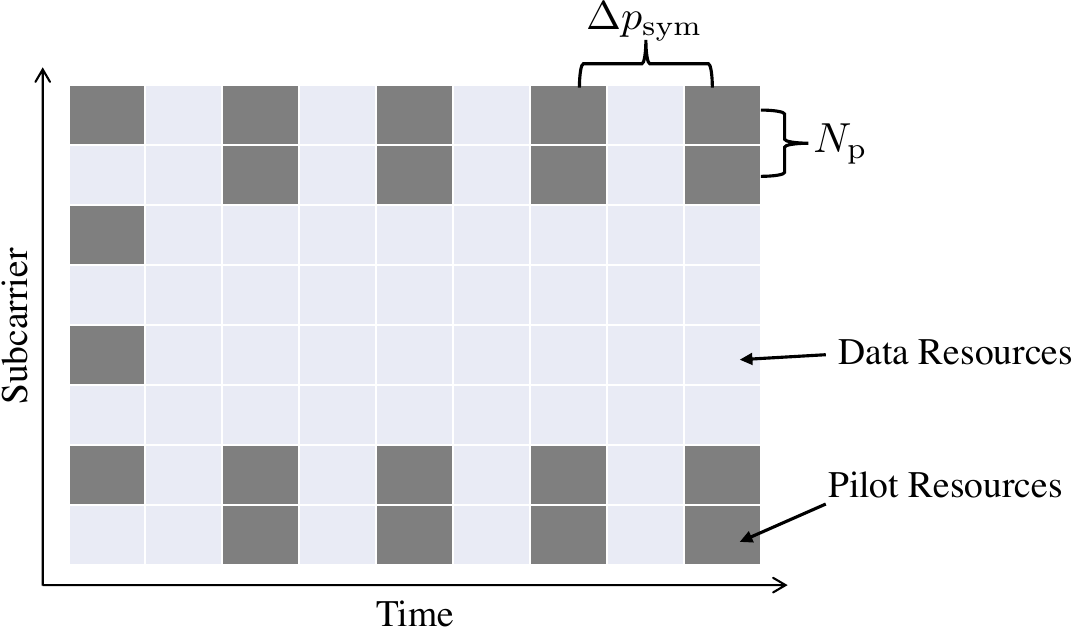}
			\caption{}
			\label{fig:signal_2}
		\end{subfigure}
		\caption{The equally-spaced-pilot signal (a) and outer-most-pilot signal (b) structures.}
	\end{figure*}

	The simulation analysis adopts an \ac{ofdm} signal structure having interspersed pilots in both time and frequency, with all other resource elements allocated for data transmission. The signal has 9 symbols in one block with $K = 72$ subcarriers and a subcarrier spacing of \SI{30}{\kilo\hertz}. The carrier frequency is $f_{\text{c}} = \SI{3.5}{\giga\hertz}$, the cyclic prefix is $L_{\text{c}} = 18$ samples, and the sampling rate is $f_{\text{s}} = \SI{2.16}{\mega\hertz}$. This results in a cyclic prefix length of \SI{8.33}{\micro\second} and a total symbol duration of \SI{41.66}{\micro\second}. The receiver has \textit{a priori} knowledge of the \ac{toa} uniformly distributed over a duration equal to the cyclic prefix length $T_{\text{a}} = \SI{8.33}{\micro\second}$. Two variants of pilot placements are explored. In the first variant, visualized in Fig.~\ref{fig:signal_1}, pilots are equally spaced in both time and frequency. The spacing over time in units of symbols is $\Delta p_{\text{sym}}$, and the spacing in frequency in units of subcarriers is $\Delta p_{\text{sc}}$. In the second variant, visualized in Fig.~\ref{fig:signal_2}, pilots are instead placed at the upper and lower subcarrier limits, with equally-spaced pilots only placed in the $m=0$ symbol spaced $2$ subcarriers apart to ensure the channel can be estimated at all subcarriers. The number of pilots placed at each extremity is $N_{\text{p}}$. The first variant will be referred to as the ``equally-spaced-pilot signal,'' while the second variant will be referred to as the ``outer-most-pilot signal.'' The equally-spaced-pilot signal is explored because it minimizes channel estimation error \cite{Ohno2004}. The outer-most-pilot signal is explored because allocating power to the extremities of the band maximizes the mean-squared-bandwidth of the signal and minimizes the \ac{crlb} on \ac{toa} estimation \cite{nanzer2016bandpass}. The equally-spaced pilots in the first symbol of the outer-most-pilot signal are included to allow accurate channel estimation across all subcarriers.
	
	The expected values over channel coefficients (i.e., in (\ref{eq:cap}), (\ref{eq:cap_mrc}), (\ref{eq:pmin_rayleigh_cdf}), and (\ref{eq:pmin_rayleigh_mrc_cdf})) and outage probabilities (i.e., in (\ref{eq:out}) and (\ref{eq:out_mrc})) are computed through Monte-Carlo methods. For the Ziv-Zakai bound, the integral in (\ref{eq:zzb}) is computed using Riemann integration sampled at step sizes of $\frac{T_{\text{s}}}{64}\;\SI{}{\second}$. Note that (\ref{eq:cap}) gives the communication capacity bound in units of \bpshz{}, and that (\ref{eq:zzb}) gives the \ac{toa} estimate variance in units of seconds-squared. To provide a more intuitive understanding of the Ziv-Zakai bound, the results plot the \ac{rmse} in units of meters: $\text{RMSE} = c\sqrt{\text{ZZB}}$, where $c$ is the speed of light in meters-per-second (m/s).
	
	To explore the impact of power allocation between pilot and data resources, define a total pilot power $P^{(m)}_{\text{pilot}}$ and total data power $P^{(m)}_{\text{data}}$. Assuming power is equally distributed among the pilot resources, $\sigma_{\tilde{s}^{(m)}}^2[k] = \frac{P^{(m)}_{\text{pilot}}}{|\mathcal{I}_{\text{pilot}}^{(m)}|}$ for $k \in \mathcal{I}_{\text{pilot}}^{(m)}$. Similarly for the data resources, $\sigma_{\tilde{s}^{(m)}}^2[k] = \frac{P^{(m)}_{\text{data}}}{|\mathcal{I}_{\text{data}}^{(m)}|}$ for $k \in \mathcal{I}_{\text{data}}^{(m)}$. The total power is equivalent over all symbols, $P^{(m)}_{\text{data}} + P^{(m)}_{\text{pilot}} = P_{\text{total}}$ for $m = 0,\; 1,\; \ldots, M-1$. Defining the set of symbols containing both pilot and data resources as $\mathcal{M} \triangleq \{m : |\mathcal{I}_{\text{pilot}}^{(m)}| \neq 0,\; |\mathcal{I}_{\text{data}}^{(m)}| \neq 0\}$, a power allocation $\alpha$  is applied such that $P^{(m)}_{\text{data}} = \alpha P_{\text{total}}$ and $P^{(m)}_{\text{pilot}} = (1-\alpha) P_{\text{total}}$ for $m \in \mathcal{M}$. The average \ac{snr} used throughout the results is defined as $\frac{P_{\text{total}} \sigma_{\tilde{h}}^2}{\sigma_{\tilde{v}}^2}$.
	
	Simulation results consider the three channel types: \ac{awgn}, frequency-flat Rayleigh fading, and frequency-selective Rayleigh fading. In both Rayleigh cases, signals are simulated with and without multiple receive antennas and \ac{mrc}. Each antenna's channel power is scaled by $\frac{1}{N_{\text{RX}}}$ to keep the total received signal power consistent, highlighting gains due to diversity rather than simply increased \ac{sinr}. Recall that the receiver knows $L$ and has statistical knowledge of the fading distributions.
	
	First, the capacity, outage, and ranging variance bounds are evaluated for the equally-spaced-pilot signal under these fading distributions, demonstrating the impact that fading has on both communications and ranging performance. Ranging variance bounds are also shown for the outer-most-pilot signal for comparison. Second, the trade-offs between communications and ranging are analyzed by plotting capacity and probability of outage against ranging error. These plots show results with different pilot resource placements in frequency and power allocations, providing insight into Pareto-optimal signal design choices. Last, capacity is plotted for varying placements of pilot resources over time to demonstrate the impact of residual phase errors.
	
	\subsection{Bounds}
	
	\begin{figure*}[h!]
		\centering
		\begin{subfigure}{0.48\linewidth}
			\centering
			\includegraphics[width=\textwidth]{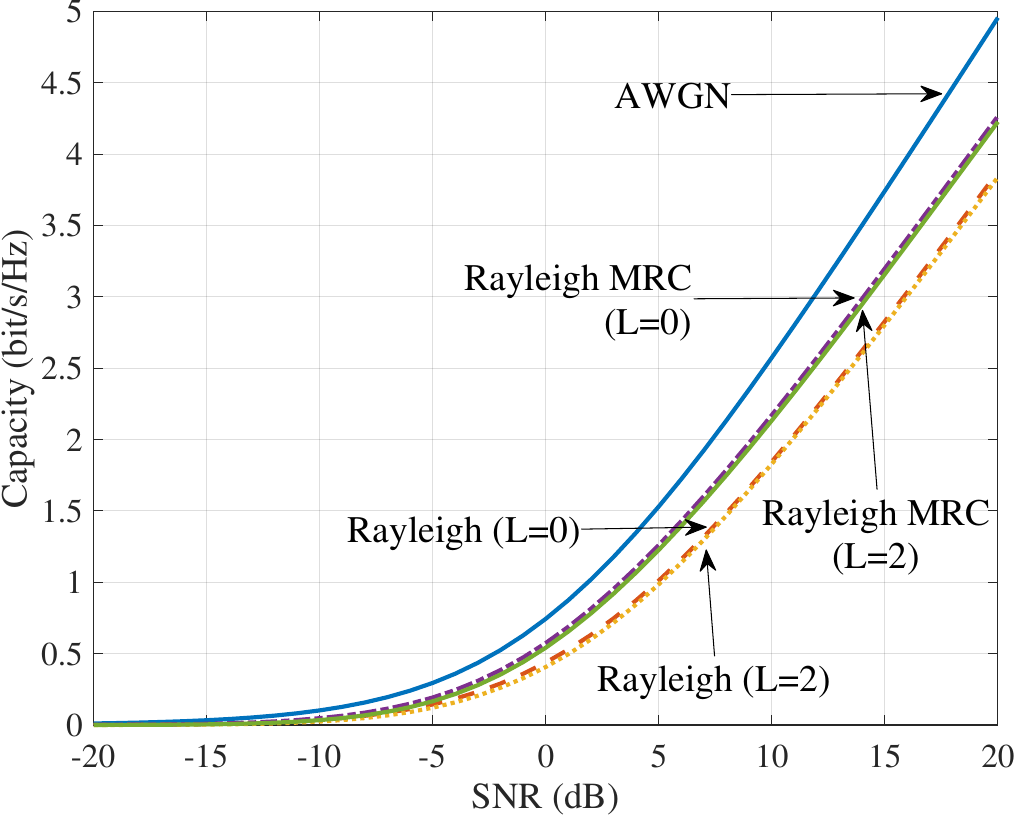}
			\caption{}
			\label{fig:cap_1}
		\end{subfigure}
		\begin{subfigure}{0.48\linewidth}
			\centering
			\includegraphics[width=\textwidth]{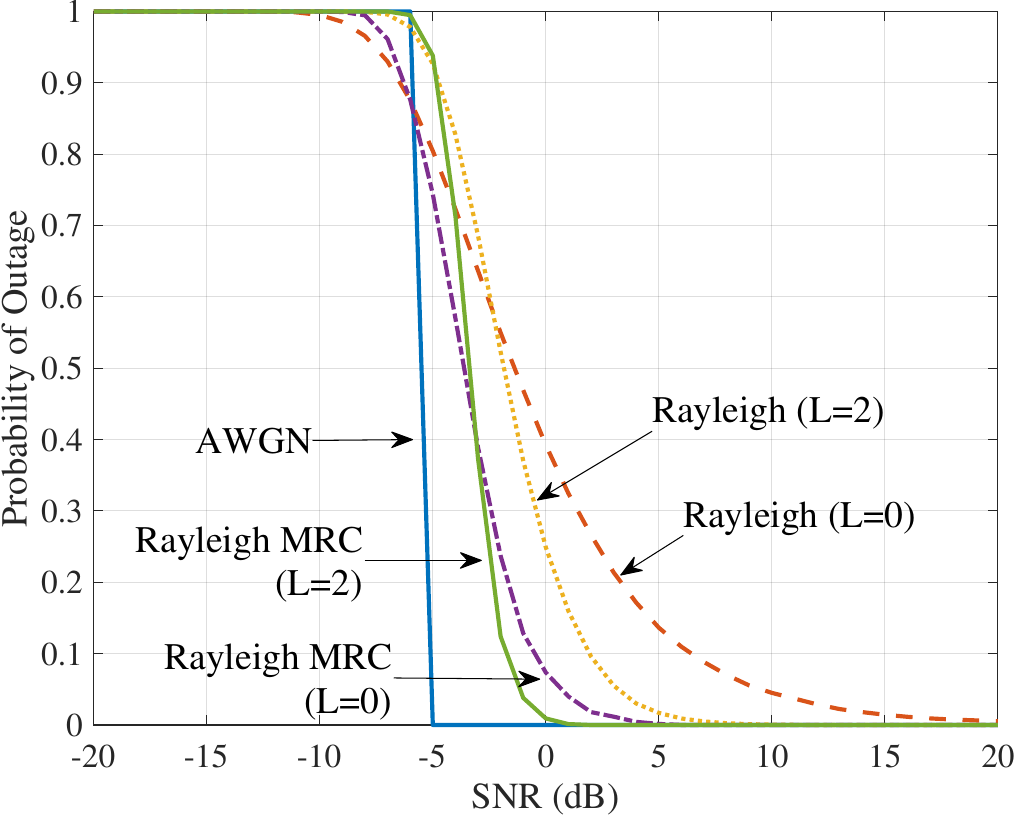}
			\caption{}
			\label{fig:out_1}
		\end{subfigure}
		\caption{The communication capacity (a) and probability of outage (b) of the equally-spaced-pilot signal as a function of \ac{snr} under different fading and channel types.}
		\label{fig:cap_out_1}
	\end{figure*}
	
	\begin{figure*}[h!]
		\centering
		\begin{subfigure}{0.48\linewidth}
			\centering
			\includegraphics[width=\textwidth]{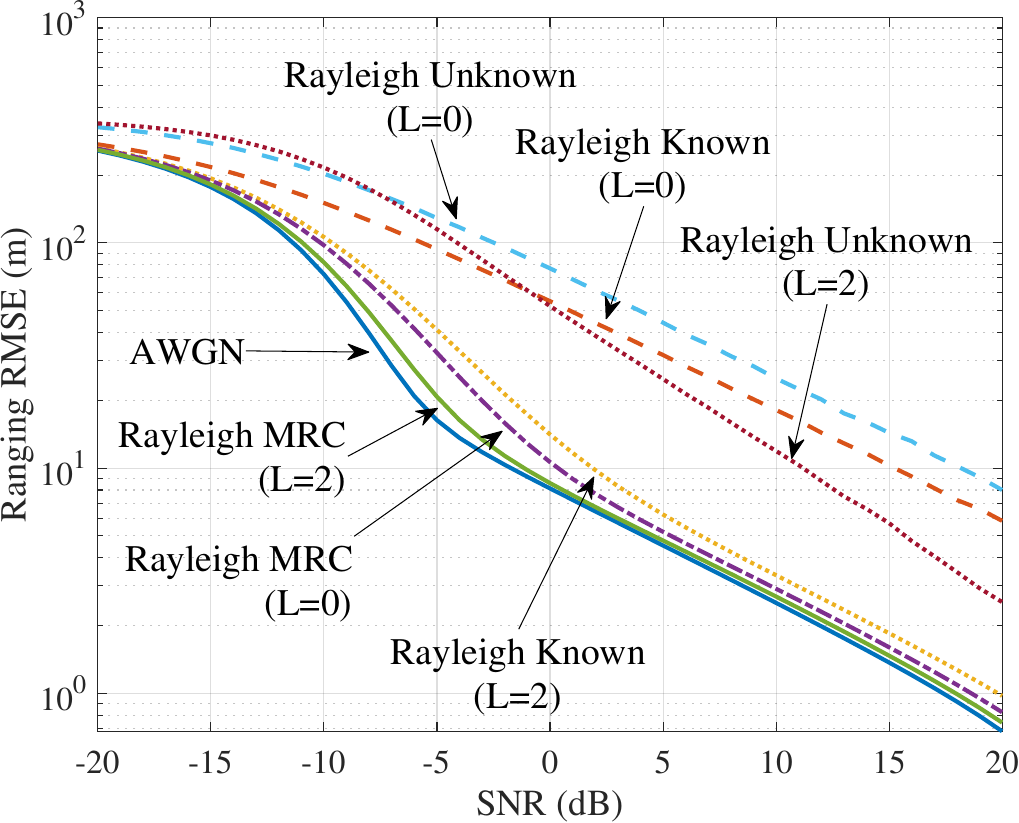}
			\caption{}
			\label{fig:zzb_1}
		\end{subfigure}
		\begin{subfigure}{0.48\linewidth}
			\centering
			\includegraphics[width=\textwidth]{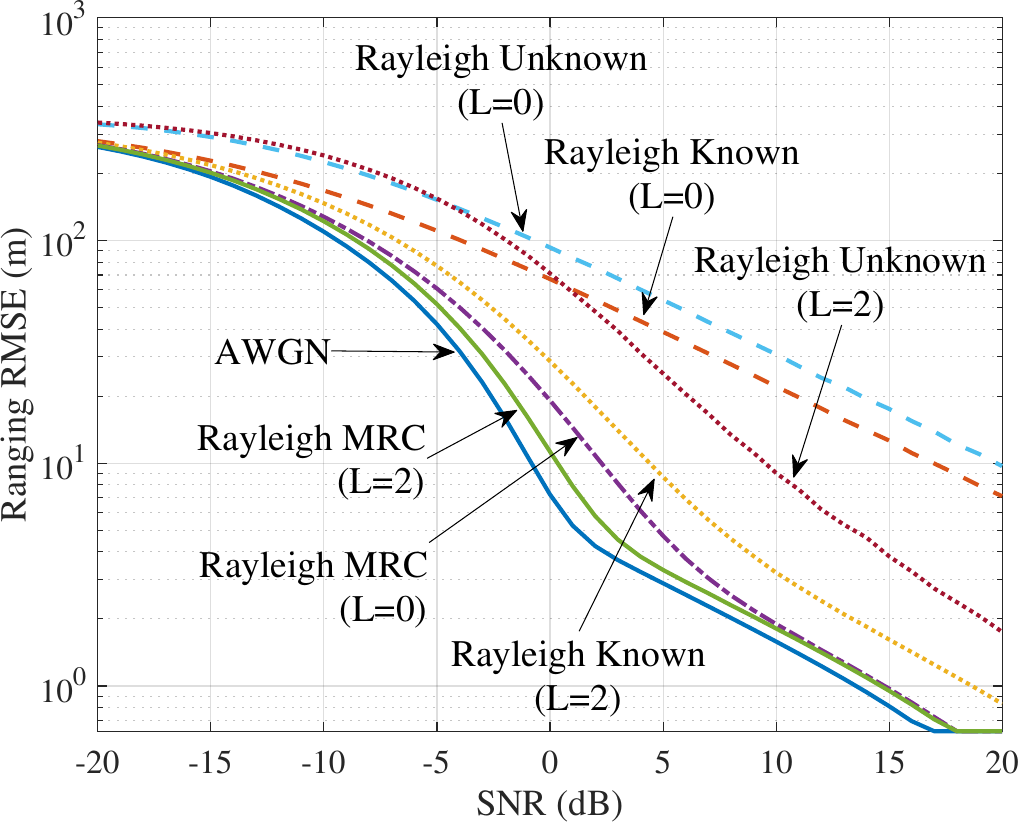}
			\caption{}
			\label{fig:zzb_2}
		\end{subfigure}
		\caption{The Ziv-Zakai Bound on the range estimate \ac{rmse} for the equally-spaced-pilot signal (a) and the outer-most-pilot signal (b) as a function of \ac{snr} under different fading and channel types.}
	\end{figure*}
	
	Capacity and outage bounds are first shown for the equally-spaced-pilot signal, with $\Delta p_{\text{sym}}=2$ and $\Delta p_{\text{sc}}=8$. Power is evenly distributed across subcarriers, resulting in the pilot resources receiving $1/8$ of the total signal power in symbols 1, 3, 5, 7, and 9. The simulation uses an outage capacity of $C_{\text{min}} = \bpshz{0.25}$, a common phase noise variance per sample of $\sigma_{\phi}^2 = \SI{0.018}{\radian^2}$, and a phase rotation limit of $\epsilon_{\phi,\text{max}} = \SI{15}{\deg}$. Fig.~\ref{fig:cap_1} shows the ergodic capacity under an \ac{awgn} channel, Rayleigh channels of orders $L=0$ and $L=2$, and Rayleigh channels of the same orders with $N_{\text{RX}}=4$ and \ac{mrc}. Fig.~\ref{fig:out_1} shows the probability of outage for the same signal and channels. Predictably, the greatest capacity is achieved under the \ac{awgn} channel, while Rayleigh fading results in reduced throughput. While increasing the order of the channel only slightly decreases capacity in the Rayleigh channels, the order $L=2$ channels exhibit reduced outages at higher \acp{snr}. When \ac{mrc} is used, the diversity gains result in an increased capacity and notably reduced probability of outage.
	
	
	Fig.~\ref{fig:zzb_1} shows the Ziv-Zakai bound for the same equally-spaced-pilot signal. In this figure, a ``known'' channel refers to bounds derived using the error probability in (\ref{eq:pmin_rayleigh_cdf}) or, when \ac{mrc} is applied, (\ref{eq:pmin_rayleigh_mrc_cdf}). The ``unknown'' channel refers to bounds derived using the error probability in (\ref{eq:pmin_cdf_none}). The \ac{awgn} channel achieves the minimum ranging \ac{rmse}. Below \SI{-5}{\decibel} \ac{snr}, the signal enters the sidelobe-dominated regime where ranging errors are exacerbated by misdetections occurring on sidelobes in the signal's ambiguity function. Above \SI{-5}{\decibel} \ac{snr}, the ranging \ac{rmse} flattens out. In the known channel case, Rayleigh fading notably degrades performance relative the \ac{awgn} channel. However, the ranging errors approach the \ac{awgn} curve as diversity is exploited through both increased channel orders and diversity gains with \ac{mrc}. The unknown Rayleigh channels result in the greatest degradation in performance at low \ac{snr}. Interestingly, the order $L=2$ unknown Rayleigh channel performs better than the order $L=0$ unknown channel above \SI{-8}{\decibel} \ac{snr} and better than the order $L=0$ known channel above \SI{-1}{\decibel} \ac{snr}.
	
	Fig.~\ref{fig:zzb_1} shows that multipath, diversity, and channel knowledge have a drastic impact on ranging \ac{rmse}. At \SI{10}{\decibel} \ac{snr}, the \ac{awgn} channel has an \ac{rmse} of \SI{2.52}{\meter}, the order $L=2$ Rayleigh \ac{mrc} channel has an \ac{rmse} of \SI{2.69}{\meter}, the order $L=0$ Rayleigh \ac{mrc} channel has an \ac{rmse} of \SI{2.91}{\meter}, and the order $L=2$ Rayleigh known channel has an \ac{rmse} of \SI{3.35}{\meter}. Significantly increasing from these values, the order $L=2$ Rayleigh unknown channel has an \ac{rmse} of \SI{11.92}{\meter}, the order $L=0$ Rayleigh known channel has an \ac{rmse} of \SI{18.12}{\meter}, and the order $L=0$ Rayleigh unknown channel has an \ac{rmse} of \SI{25.02}{\meter}. 
	
	In comparison, Fig.~\ref{fig:zzb_2} shows the Ziv-Zakai bound for the outer-most-pilot signal, exhibiting structure similar to Fig.~\ref{fig:zzb_1}. The mainlobe-dominated regime of this signal achieves smaller errors than the equally-spaced-pilot signal. For example, at \SI{10}{\decibel} \ac{snr}, the \ac{awgn} \ac{rmse} is reduced from \SI{2.52}{\meter} to \SI{1.58}{\meter}. The outer-most-pilot signal achieves a lower \ac{rmse} than the equally-spaced-pilot signal above \SI{0}{\decibel} in the \ac{awgn} channel, above \SI{1}{\decibel} in the order $L=2$ Rayleigh \ac{mrc} channel, above \SI{4}{\decibel} in the order $L=0$ Rayleigh \ac{mrc} channel, above \SI{9}{\decibel} in the order $L=2$ Rayleigh known channel, and above \SI{6}{\decibel} in the order $L=2$ Rayleigh unknown channel. For all \ac{snr} below these thresholds and for all other channels, the outer-most-pilot signal performs worse than the equally-spaced-pilot signal, due to the different shapes of each signal's autocorrelation function. The allocation of power to the extremities in the outer-most-pilot signal sharpens the mainlobe of the autocorrelation function at the expense of higher sidelobes, resulting in reduced ranging errors only when the \ac{snr} remains high enough to prevent detections on the sidelobes. Thus, this paper's bounds offer an immediate and valuable insight: allocating pilot resources to the subcarrier extremities reduces ranging errors in high \ac{snr} channels when \ac{mrc} and \textit{a priori} multipath knowledge can be exploited, but an equally-spaced pilot allocation achieves better performance at lower \acp{snr} and when fading is frequency-flat.
	
	\subsection{Pareto Curves}
	
	\begin{figure*}[h!]
		\centering
		\begin{subfigure}{0.48\linewidth}
			\centering
			\includegraphics[width=\textwidth]{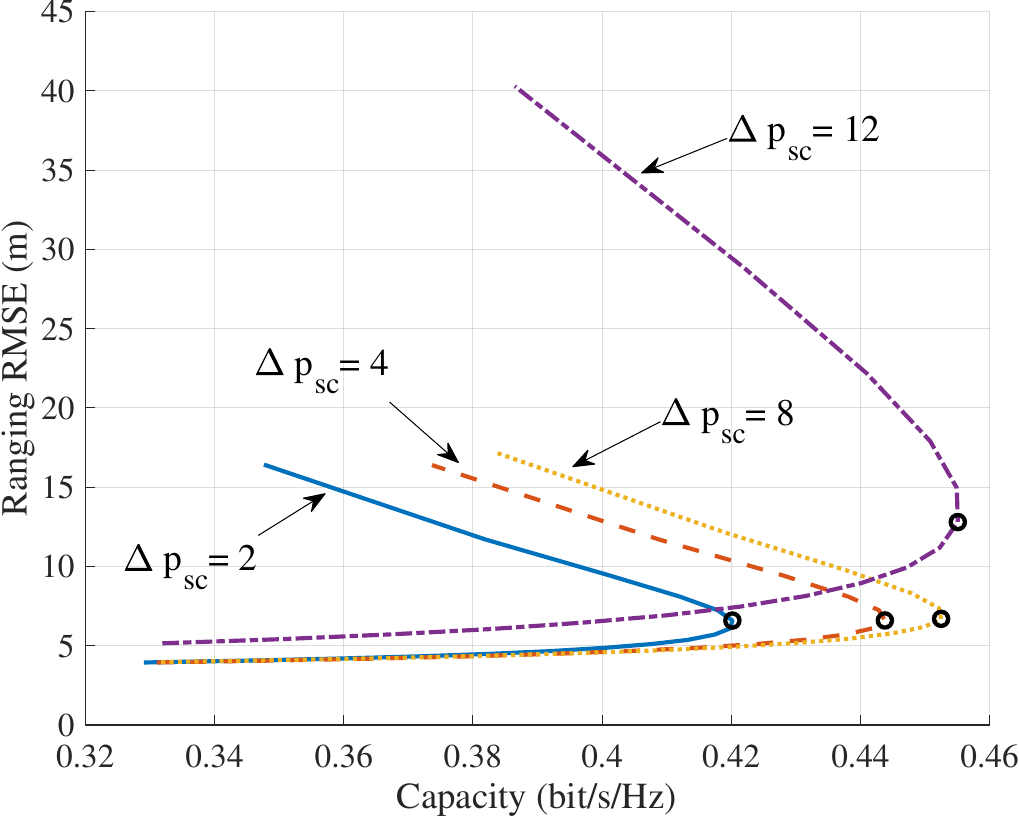}
			\caption{}
			\label{fig:pareto_cap_1}
		\end{subfigure}
		\begin{subfigure}{0.48\linewidth}
			\centering
			\includegraphics[width=\textwidth]{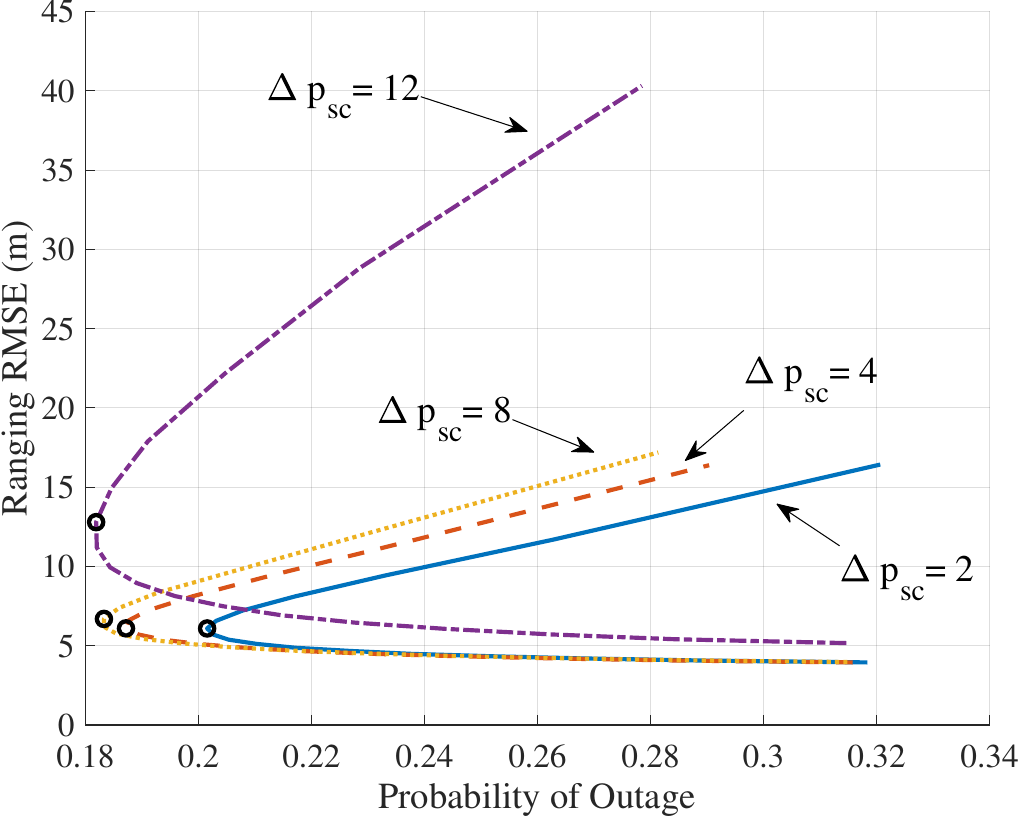}
			\caption{}
			\label{fig:pareto_out_1}
		\end{subfigure}
		\caption{Pareto trade-off curves plotting capacity (a) and probability of outage (b) against ranging \ac{rmse} for the equally-spaced-pilot signal at \SI{0}{\decibel} \ac{snr}. Translation upward along each curve corresponds to increasing  $\alpha$, the fraction of power allocated to the data resources when pilots are present. The black dots correspond to values of $\alpha$ that maximize the capacity. Curves are plotted for multiple pilot resource spacings across the subcarriers, $\Delta p_{\text{sc}}$. The channel is order $L=2$ Rayleigh, and channel coefficients are assumed known for ranging only.}
		\label{fig:pareto_1}
	\end{figure*}
	
	Now, the tradeoff between ranging and communications performance is analyzed for both signal structures. Pareto curves are plotted between capacity and ranging \ac{rmse} as well as outage probability and ranging \ac{rmse}. Each simulation uses an \ac{snr} of \SI{0}{\decibel}, an outage capacity of $C_{\text{min}} = \bpshz{0.25}$, a common phase noise variance per sample of $\sigma_{\phi}^2 = \SI{0.018}{\radian^2}$, and a phase rotation limit of $\epsilon_{\phi,\text{max}} = \SI{15}{\deg}$. The data power fraction $\alpha$ is varied from 0.1 to 0.9.
	
	Figs.~\ref{fig:pareto_cap_1} and \ref{fig:pareto_out_1} show the capacity and outage Pareto curves for the equally-spaced-pilot signal under an order $L=2$ Rayleigh channel. For the ZZB, perfect channel knowledge is assumed. The pilot spacing across symbols is fixed at $\Delta p_{\text{sym}} = 2$, and each curve corresponds to a different pilot spacing across subcarriers, $\Delta p_{\text{sc}}$. The maximum capacity of this system is achieved at $\Delta p_{\text{sc}} = 12$ and $\alpha \approx 0.65$. As more power is allocated to the pilot resources by decreasing $\alpha$, ranging \ac{rmse} is reduced by moving downward along the $\Delta p_{\text{sc}} = 12$ curve at the expense of data throughput. However, it becomes advantageous to use a pilot spacing of $\Delta p_{\text{sc}} = 8$ to continue maximizing capacity if a reduced ranging \ac{rmse} is desired. In fact, the capacity-maximizing power allocation for $\Delta p_{\text{sc}} = 8$ approximately halves the ranging \ac{rmse} compared to $\Delta p_{\text{sc}} = 12$ while only sacrificing \bpshz{0.003}. Interestingly, pilot spacings of $\Delta p_{\text{sc}} = 4$ and $\Delta p_{\text{sc}} = 2$ are never Pareto-optimal for $0.1 \leq \alpha \leq 0.9$. Similar results are seen in the outage Pareto curve. One explanation for this behavior is that increasing $\Delta p_{\text{sc}}$ allows more resources to be allocated for communications, increasing capacity so long as power is allocated appropriately to handle channel and phase estimation errors. However, increasing this spacing also changes the shape and power of the sidelobes in the autocorrelation function, potentially increasing ranging errors.
	
	\begin{figure*}[h!]
		\centering
		\begin{subfigure}{0.48\linewidth}
			\centering
			\includegraphics[width=\textwidth]{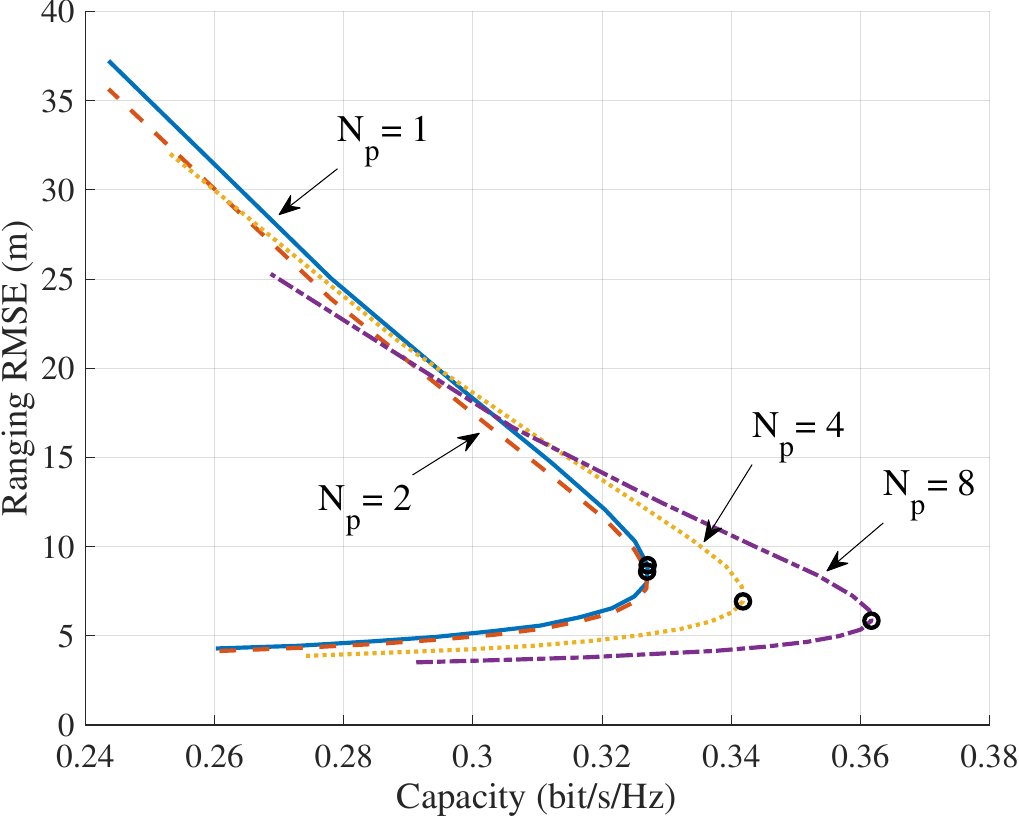}
			\caption{}
			\label{fig:pareto_cap_2}
		\end{subfigure}
		\begin{subfigure}{0.48\linewidth}
			\centering
			\includegraphics[width=\textwidth]{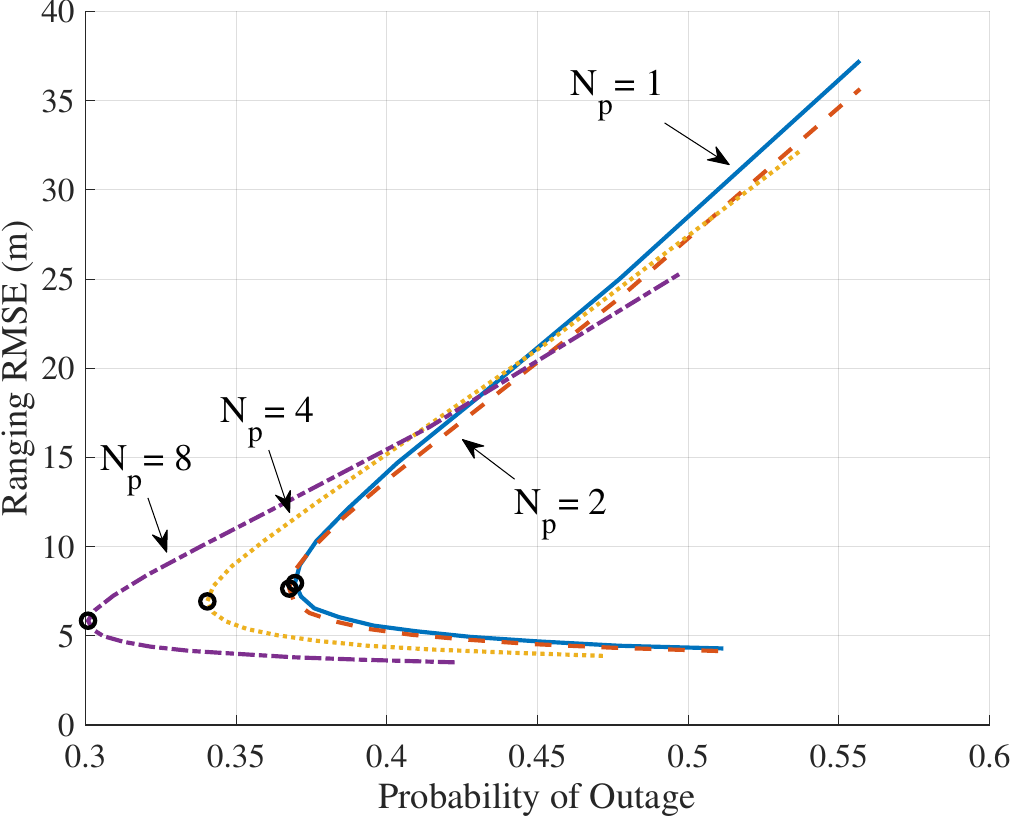}
			\caption{}
			\label{fig:pareto_out_2}
		\end{subfigure}
		\caption{As Fig.~\ref{fig:pareto_1} but for the outer-most-pilot signal.}
	\end{figure*}
	
	Figs.~\ref{fig:pareto_cap_2} and \ref{fig:pareto_out_2} show the capacity and outage Pareto curves for the same simulation setup but using the outer-most-pilot signal. This signal structure results in a different shape for the Pareto-optimal design where $N_{\text{p}}=8$ is the Pareto-optimal choice when power is allocated appropriately. Along the $N_{\text{p}}=8$ curve below \bpshz{0.33}, slightly lower ranging \ac{rmse}s are achieved compared to the equally-spaced-pilot signal. However, the capacity-maximizing power allocation in this signal structure loses approximately \bpshz{0.09} compared to the maximum capacity of the equally-spaced-pilot signal. Similar results are seen in the probability of outage, where the outer-most-pilot signal cannot minimize outages as well as the equally-spaced-pilot signal but can achieve slightly reduced ranging \ac{rmse} if the increased outages are tolerable. This behavior may be caused by increased channel estimation errors since pilots are not distributed equally across the band in the outer-most-pilot signal.
	
	\begin{figure*}[h!]
		\centering
		\begin{subfigure}{0.48\linewidth}
			\centering
			\includegraphics[width=\textwidth]{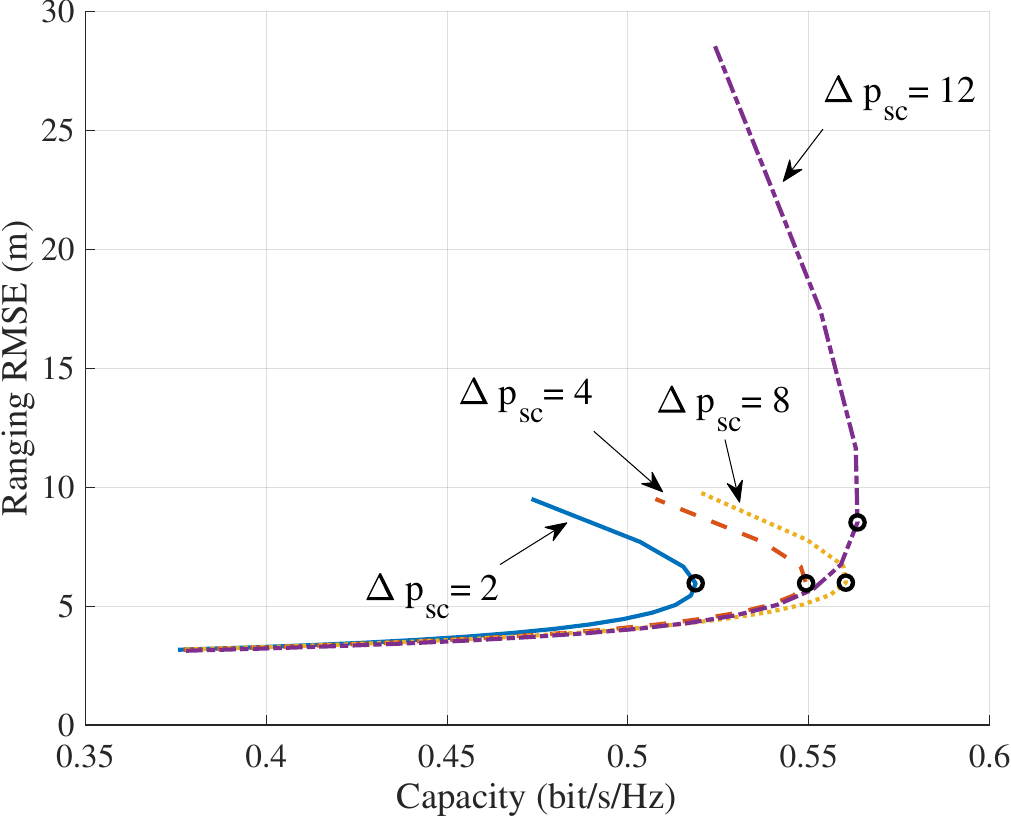}
			\caption{}
			\label{fig:pareto_cap_3}
		\end{subfigure}
		\begin{subfigure}{0.48\linewidth}
			\centering
			\includegraphics[width=\textwidth]{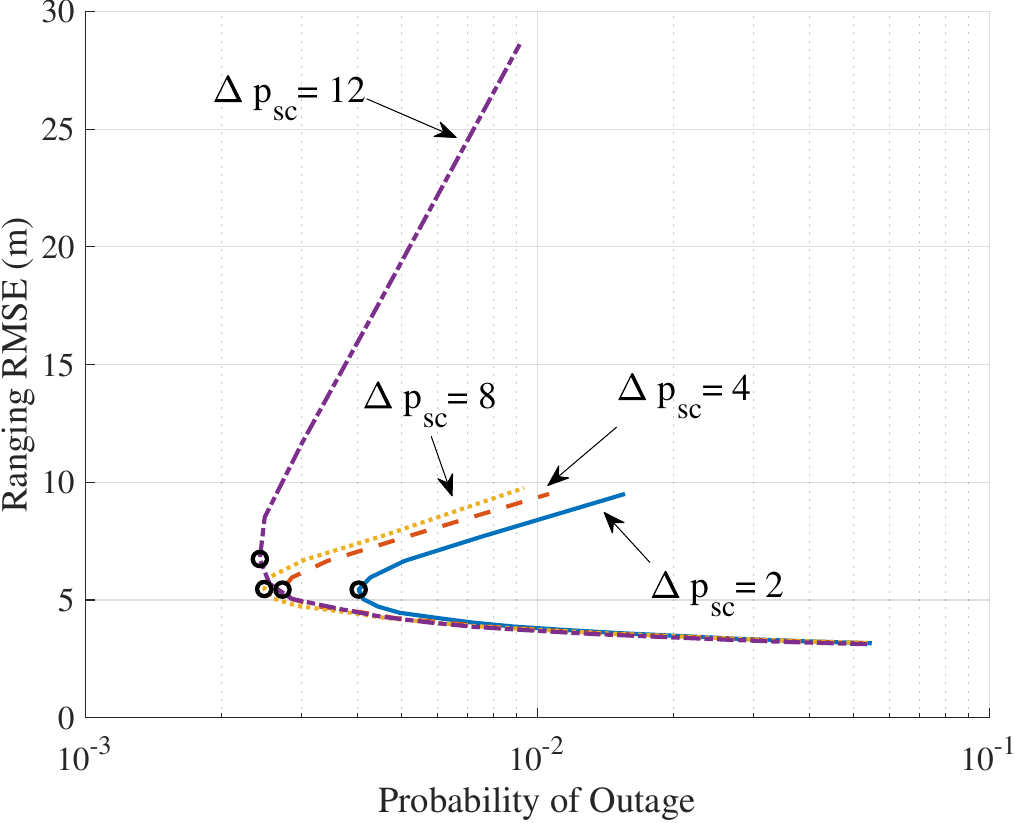}
			\caption{}
			\label{fig:pareto_out_3}
		\end{subfigure}
		\caption{Pareto trade-off curves plotting capacity (a) and outage (b) against ranging \ac{rmse} for the equally-spaced-pilot signal at \SI{0}{\decibel} \ac{snr}. The channel is order $L=2$ Rayleigh and \ac{mrc} with $N_{\text{RX}} = 4$ is used.}
	\end{figure*}
	
	This same analysis can be repeated with an order $L=2$ Rayleigh channel with \ac{mrc}. As seen in the previous analysis, \ac{mrc} provides immense benefits to capacity, outage, and ranging \ac{rmse}. Figs.~\ref{fig:pareto_cap_3} and \ref{fig:pareto_out_3} show the capacity and outage Pareto curves for the equally-spaced-pilot signal with $N_{\text{RX}} = 4$ and \ac{mrc}. Compared to Fig.~\ref{fig:pareto_cap_1}, the Pareto-optimal design with \ac{mrc} achieves greater capacity and reduced ranging \ac{rmse} for all power allocations. More notable, however, is the reduction in outages that \ac{mrc} provides, reducing outage probability down to a minimum of \SI{2.4e-3}{}. \ac{mrc} is able to achieve the results by decreasing the likelihood of experiencing deep fades and extremely low \acp{snr} by exploiting antenna diversity.
	
	\begin{figure*}[h!]
		\centering
		\begin{subfigure}{0.48\linewidth}
			\centering
			\includegraphics[width=\textwidth]{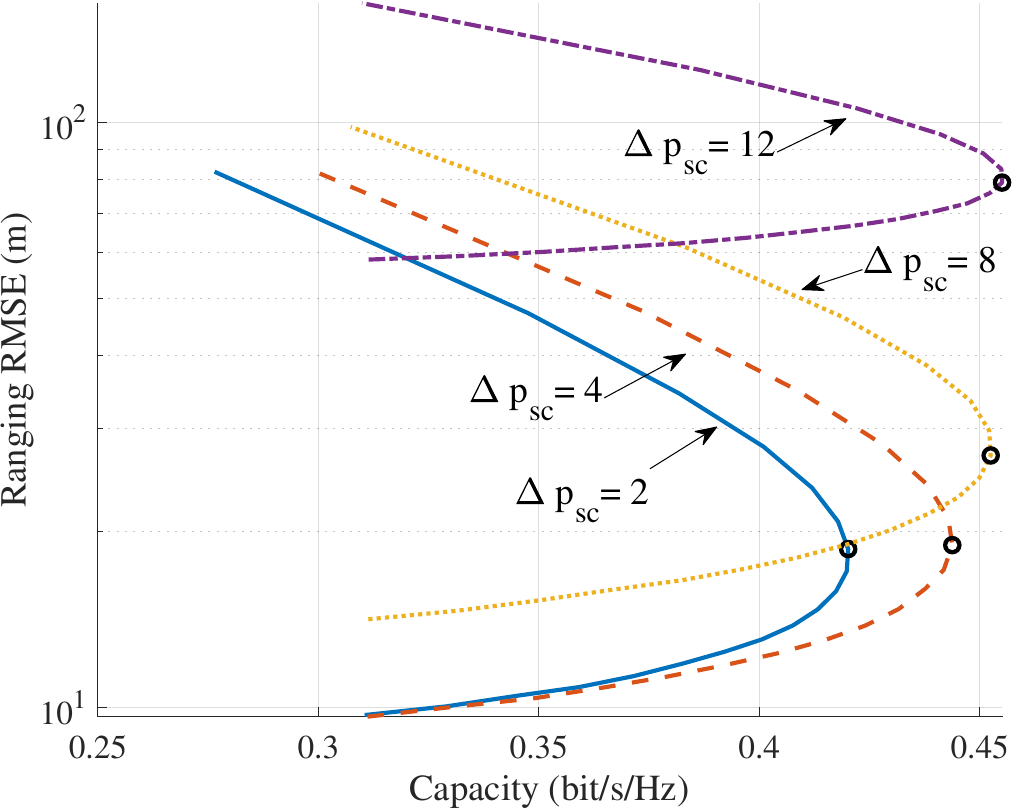}
			\caption{}
			\label{fig:pareto_cap_4}
		\end{subfigure}
		\begin{subfigure}{0.48\linewidth}
			\centering
			\includegraphics[width=\textwidth]{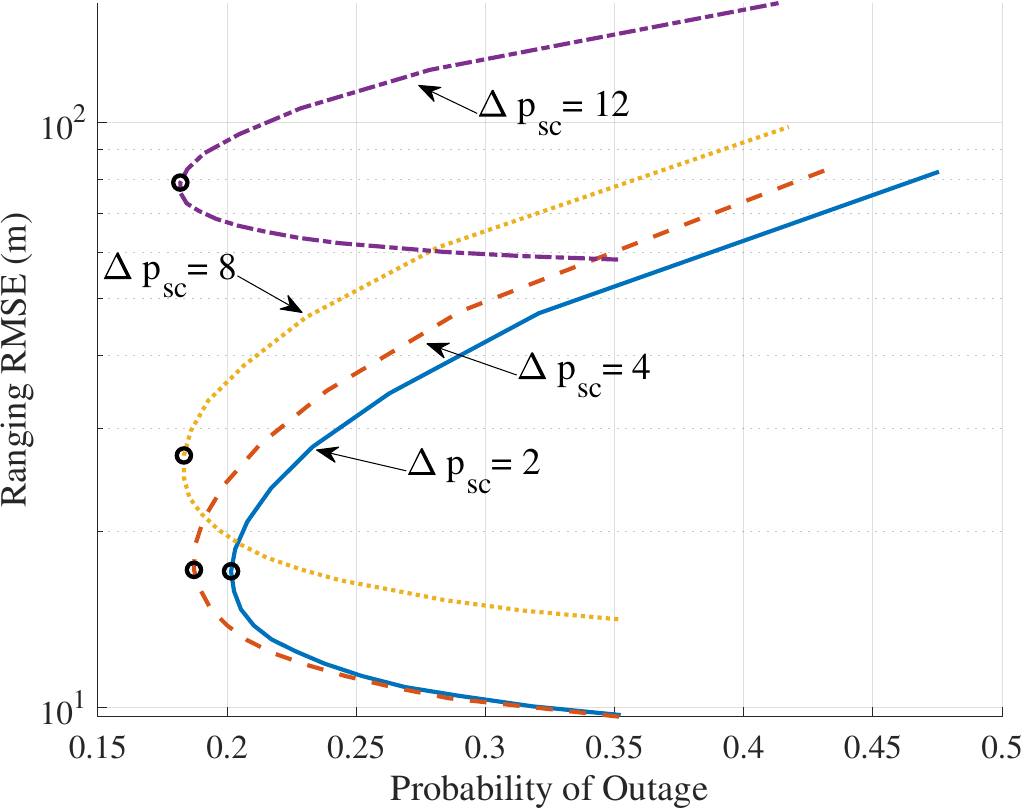}
			\caption{}
			\label{fig:pareto_out_4}
		\end{subfigure}
		\caption{Pareto trade-off curves plotting capacity (a) and outage (b) against ranging \ac{rmse} for the equally-spaced-pilot signal at \SI{0}{\decibel} \ac{snr} in an order $L=2$ unknown Rayleigh channel.}
	\end{figure*}
	
	The last Pareto plots in Figs.~\ref{fig:pareto_cap_4} and \ref{fig:pareto_out_4} show the capacity and outage for the equally-spaced-pilot signal in an order $L=2$ unknown Rayleigh channel. Much like Figs.~\ref{fig:pareto_cap_1} and \ref{fig:pareto_out_1}, capacity is maximized on the $\Delta p_{\text{sc}} = 12$ curve. However, significant improvements in ranging accuracy can be gained if the system instead opts for a pilot resource spacing of $\Delta p_{\text{sc}} = 8$ or $\Delta p_{\text{sc}} = 4$, reducing the ranging \ac{rmse} at the capacity-maximizing point from \SI{79.0}{\meter} to \SI{27.0}{\meter} at the expense of \bpshz{0.003}, or to \SI{19.0}{\meter} at the expense of \bpshz{0.110}. These poor ranging results are caused by an inability of the receiver to exploit multipath diversity when the channel is unknown, making it susceptible to deep fades that drastically increase ranging errors.
	
	\begin{figure}[h!]
		\centering
		\includegraphics[width=0.48\textwidth]{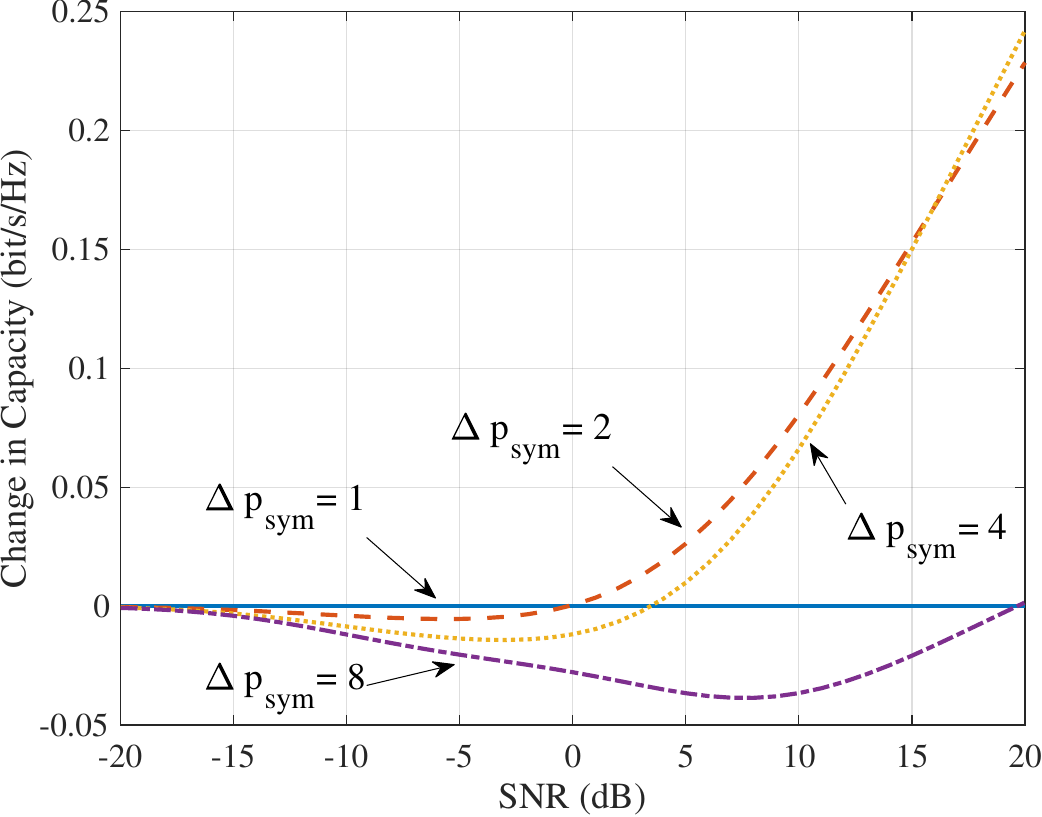}
		\caption{Change in communication capacity as a function of \ac{snr} for the equally-spaced-pilot signal with $\Delta p_{\text{sc}} = 8$ as $\Delta p_{\text{sym}}$ is varied. The capacity achieved with $\Delta p_{\text{sym}} = 1$ is used as the baseline. The channel is order $L=2$ Rayleigh.}
		\label{fig:sym_sp_cap_rel_1}
	\end{figure}
	
	Finally, the impact of pilot resource spacing across symbols is analyzed for the equally-spaced-pilot signal. The spacing $\Delta p_{\text{sym}}$ is varied between 1, 2, 4, and 8. The channel is order $L=2$ Rayleigh and the pilot resource spacing in frequency is $\Delta p_{\text{sc}} = 8$. The relative capacity curves in Fig.~\ref{fig:sym_sp_cap_rel_1} show that $\Delta p_{\text{sym}} = 1$ maximizes capacity for \acp{snr} below \SI{0}{\decibel}, whereas an increased $\Delta p_{\text{sym}}$ reduces \ac{cfo}, \ac{cpe}, and channel estimation accuracy, degrading the throughput. Even at high \acp{snr} where channel estimation errors are small, uncompensated phase errors increase when $\Delta p_{\text{sym}}$ is too large, reducing capacity because of the phase error requirement imposed in (\ref{eq:rate}). However, $\Delta p_{\text{sym}} = 2$ maximizes capacity from \SI{0}{\decibel} to \SI{16}{\decibel} \ac{snr}, and $\Delta p_{\text{sym}} = 4$ maximizes capacity above \SI{16}{\decibel} \ac{snr}. With this information, a network can optimize the spacing of pilot resources in time to maximize capacity for a user given knowledge of its receiver's \ac{snr}.
	
	\section{Conclusion}
	This paper has demonstrated how \ac{ofdm} pilot resource allocations can be analyzed and designed for both ranging purposes and communications. Bounds were derived to capture the impact that the placement and power allocation of pilot resources have on communications capacity, outages, and the Ziv-Zakai bound on ranging variance. These bounds highlighted how multipath and receive diversity can reduce outages and improve ranging accuracy. Furthermore, ranging accuracy is significantly impacted by the receiver's existing knowledge of the channel. Pareto-optimal pilot resource designs were explored, revealing that ranging errors can be improved with minimal degradation in communications throughput. It was further shown that the spacing of pilots in time can be adjusted to maximize capacity over a wide range of \acp{snr}.
	
	As users continue to demand precise positioning from communication networks, designers of next-generation protocols will need to rethink basic \ac{ofdm} design. The results in this paper demonstrate that \ac{ofdm} signals can be designed in a manner that balances the tradeoffs between ranging and communications.
	
	\nolinenumbers
	
	\renewcommand{\IEEEiedlistdecl}{\IEEEsetlabelwidth{SONET}}
	\printacronyms[name=Abbreviations]
	\renewcommand{\IEEEiedlistdecl}{\relax}
	
	\section*{Declarations}
	
	\subsection*{Availability of data and materials}
	
	The datasets used and/or analyzed during the current study are available from the corresponding author on reasonable request.
	
	\subsection*{Competing interests}
	
	The authors declare that they have no competing interests.
	
	\subsection*{Funding}
	Research was supported by the U.S. Department of Transportation under Grant
	69A3552348327 for the CARMEN+ University Transportation Center, and by Keysight,
	an affiliate of the 6G@UT center within the Wireless Networking and
	Communications Group at The University of Texas at Austin.
	
	\subsection*{Author contributions}
	
	The authors equally contributed to the paper. AG wrote the manuscript, conducted the simulations, and generated the results. TH provided supervision and edited the manuscript.
	
	\subsection*{Acknowledgements}
	
	Not applicable.
	
	\appendix
	
	\label{app:pn}
	
	This appendix derives the covariance for correlated noise under the approximation in \cite{Tretter1985}. Assume a signal with amplitude $A[k]$ and additive noise $v[k]$, denoted in vector form as $\bm{v} \sim \mathcal{C}\mathcal{N}(\bm{0},\bm{\Sigma})$. The additive noise is approximated as phase noise $v_{\phi}[k] = \mathfrak{I}\left(\frac{1}{A[k]} {v}[k]\right)$. It follows that the expected value is $\mathbb{E}[v_{\phi}[k]] = 0$, and the covariance is $\mathbb{E}[v_{\phi}[k]v_{\phi}[l]] = \frac{\mathfrak{R}\left((\bm{\Sigma})_{kl}\right)}{2A[k]A[l]}$. The values from the paper can be substituted in as $\bm{v} = \tilde{\bm{v}}^{(m)}_{\text{total}}$, $\bm{\Sigma} = \bm{\Sigma}^{(m)}_{\tilde{v}_{\text{total}}}$, and $A[k] = \sqrt{P^{(m)}_k P_{\text{att}} \gamma_k}$. As a result, $\tilde{\bm{v}}^{(m)}_{\text{total}}$ can be approximated as phase noise $\tilde{\bm{v}}^{(m)}_{\phi,\text{total}} \sim \mathcal{N}\left( \bm{0}, \bm{\Sigma}^{(m)}_{\phi} \right)$, where $\left(\bm{\Sigma}^{(m)}_{\phi}\right)_{kl} = \frac{\mathfrak{R}\left(\left(\bm{\Sigma}^{(m)}_{\tilde{v}_{\text{total}}}\right)_{kl}\right)}{2 P_{\text{att}} \sqrt{P^{(m)}_k\gamma_k}\sqrt{P^{(m)}_l \gamma_l}}$.

	\clearpage
	\bibliographystyle{IEEEtran}
	\bibliography{./pangea}
	
\end{document}